\documentclass[11pt,fleqn]{article}

\usepackage{amsmath,amssymb,amsthm,enumerate,cite}

\newcommand{\p}{\partial}
\newcommand{\const}{\mathop{\rm const}\nolimits}

\newcommand{\sign}{\mathop{\rm sign}\nolimits}
\newcommand{\pr}{\mathop{\rm pr}\nolimits}
\newcommand{\thetbn}{\arabic{nomer}}
\newcommand{\CV}{\mathop{\rm CV}\nolimits}
\newcommand{\CL}{\mathop{\rm CL}\nolimits}
\newcommand{\Ch}{\mathop{\rm Ch}\nolimits}

\newcounter{tbn}
\newcounter{casetran}
\newcounter{clnumber}

\newtheorem{theorem}{Theorem}

\newtheorem{proposition}{Proposition}
{\theoremstyle{definition} \newtheorem{definition}{Definition}

\newtheorem{remark}{Remark}

\begin{document}

\par\noindent {\LARGE\bf Group analysis and exact solutions of a class\\
of variable coefficient nonlinear telegraph equations\par}
{\vspace{4mm}\par\noindent {\bf  Ding-jiang Huang~$^\dag$ and Nataliya~M.~Ivanova~$^\ddag$
} \par\vspace{2mm}\par}
{\vspace{2mm}\par\noindent {\it
$^\dag$~Department of Applied Mathematics, Dalian University of Technology,  Dalian,\\
$\phantom{^\dag}$~116024, P. R. China\\
}}
{\noindent \vspace{2mm}{\it
$\phantom{^\dag}$~e-mail: hdj8116@163.com
}\par}

{\par\noindent\vspace{2mm} {\it
$^\ddag$~Institute of Mathematics of NAS of Ukraine, 
3 Tereshchenkivska Str., 01601 Kyiv, Ukraine\\[-1ex]
}}
{\noindent {\it
$\phantom{^\S}$~e-mail: ivanova@imath.kiev.ua
} \par}

{\vspace{5mm}\par\noindent\hspace*{8mm}\parbox{140mm}{\small
A complete group
classification of a class of variable coefficient
(1+1)-dimensional telegraph equations
$f(x)u_{tt}=(H(u)u_x)_x+K(u)u_x$, is given, by using a
compatibility method and additional equivalence transformations. A
number of new interesting nonlinear invariant models which have
non-trivial invariance algebras are obtained.
Furthermore, the possible additional equivalence transformations between equations from the
class under consideration are investigated. Exact solutions of
special forms of these equations are also constructed
via classical Lie method and generalized conditional transformations.
Local conservation laws with characteristics of order 0 of the class under consideration
are classified with respect to the group of equivalence transformations.
}\par\vspace{5mm}}

\section{Introduction}

Since Sophus Lie (1842--1899) introduced the notion of
continuous transformation group, now known as Lie group, the
theory of Lie groups and Lie algebras have been evolved into one of
the most explosive development of mathematics and physics
throughout the past century. Nowadays, this theory has been
widely applied to diverse fields of mathematics including
differential geometry, algebraic topology, bifurcation theory,
numerical analysis, special functions and to nearly any area of
theoretical physics, in particular classical, continuum and
quantum mechanics, fluid dynamics system, relativity, and particle
physics~\cite{Bluman&Anco2002,Bluman&Kumei1989,
Fushchych&Shtelen&Serov1993,Ibragimov1985,Ibragimov1994V1,
Ibragimov1999,Olver1986,Ovsiannikov1982,Stephani1994}.

When applied to system of differential equations, the mathematical
trends whose object is a common treatment of the Lie groups of
transformations and the differential equations admitted by these
groups is called group theoretical analysis of differential
equation. Traditionally, there are two interrelated problems in
this subject. The first one is finding the maximal Lie (symmetry)
transformation group admitted by a given equation. The
second problem is classifying differential equations that admit a
prescribed symmetry group~$G$. The principal tool for handling
both problems is the classical infinitesimal routine developed by
S.~Lie (see, e.g.,~\cite{Olver1986,Ovsiannikov1982}).
It reduces the problem to finding the
corresponding Lie symmetry algebra of infinitesimal operators
whose coefficients are found as solutions of some over-determined
system of linear partial differential equations.

The problems of group classification and exhaustive solutions of
such problems are not only interesting from the purely
mathematical point of view, but also important for applications.
It is well known that modelling the phenomena in nature (such as
in physics, chemistry and biology) by partial differential
equations is one of the central problems of mathematical physics
and applied mathematics. Generally, those modelling differential
equations could contain some arbitrary parameters or functions
which have been found experimentally and so are not strictly
fixed. Therefore, in order to reflect the natural laws accurately
one has to decide which differential equation fits in the best way
as a model for the process under study and so has to select from a
broad class of possible partial differential equations. Because in
many physical models there often exist a priori requirements for
symmetry groups that follow from physical laws (in particular,
from Galilean or relativistic theory), which imply that solving
the problems of group classification makes it possible to accept
for the criterion of applicability the following statement:
modelling differential equations have to admit a group with
certain properties or the most extensive symmetry group from the
possible ones. This point of view is supported by the fact that
the most successful mathematical models in theoretical and applied
science have a rich symmetry structure. Indeed, the basic
equations of modern physics, the wave, Schr\"{o}dinger, Dirac and
Maxwell equations are distinguished from the whole set of partial
differential equations by their Lie and non-Lie (hidden) symmetries
(see, e.g.,~\cite{Fushchych&Shtelen&Serov1993}
for more details on symmetry properties of these
equations).

In the approach used here, an exhaustive consideration of the problem of group classification
for a parametric class $\mathcal{L}$ of systems of differential equations includes the following steps:
\begin{enumerate}\itemsep=0ex
\item
Finding the group $G^{\ker}$ (the kernel of maximal Lie invariance groups)
of local transformations that are symmetries for all systems from $\mathcal{L}$.
\item
Construction of the group $G^{\sim}$ (the equivalence group) of local transformations
which transform $\mathcal{L}$ into itself.
\item
Description of all possible $G^{\sim}$-inequivalent values of parameters that admit
maximal invariance groups wider than $G^{\ker}$.
\end{enumerate}
Following S.~Lie, one usually considers infinitesimal transformations instead of finite ones.
This approach essentially simplifies the problem of group classification,
reducing it to problems for Lie algebras of vector fields.
See~\cite{Nikitin&Popovych2001,Zhdanov&Lahno1999,Ovsiannikov1982,
Popovych&Ivanova2004NVCDCEs,Akhatov&Gazizov&Ibragimov1987,Akhatov&Gazizov&Ibragimov1989}
for precise formulation of group classification problems
and more details on the used methods.

The result of application of the above algorithm is a list of
equations with their Lie invariance algebras. The problem of group
classification is assumed to be completely solved if

\begin{enumerate}[\it i\rm)]
\item
the list contains all the possible inequivalent cases of extensions;
\item\vspace{-1ex}
all the equations from the list are mutually inequivalent with respect to
the transformations from $G^{\sim}$;
\item\vspace{-1ex}
the obtained algebras are the maximal invariance algebras of the respective equations.
\end{enumerate}

Such a list may include equations that are mutually equivalent
with respect to local transformations which do not belong to
$G^{\sim}$. Knowing such additional equivalences allows one to
essentially simplify further investigation of $\mathcal{L}$.
Constructing them can be considered as the fourth step of the
algorithm of group classification. Then, the above enumeration of
requirements for the resulting list of classifications can be
completed by the following step:

\begin{enumerate}[\it i\rm)]\setcounter{enumi}{3}\itemsep=0ex
\item
all the possible additional equivalences between the listed equations are constructed in
explicit form.
\end{enumerate}

In this paper we consider a class of variable coefficient
(1+1)-dimensional nonlinear telegraph equations of the form
\begin{equation}\label{eqVarCoefTelegraphEq}
f(x)u_{tt}=(H(u)u_x)_x+K(u)u_x
\end{equation}
where $f=f(x)$, $H=H(u)$ and $K=K(u)$ are arbitrary and sufficient
smooth real-valued function of their corresponding variable,
$f(x)H(u)\neq 0$. In what follows, we assume that $(H_u, K_u)\neq(0, 0)$,
i.e.,~\eqref{eqVarCoefTelegraphEq} is a nonlinear equation. This is because the
linear case of~\eqref{eqVarCoefTelegraphEq} ($H, K=\const$) has been studied by Lie~\cite{Lie1881} in his
classification of linear second-order PDEs with two variables.
(See also a modern treatment of this subject in~\cite{Ovsiannikov1982}).

The study of equation~\eqref{eqVarCoefTelegraphEq} is stimulated not only their intrinsic
theoretical interest, but also its physical importance.
Equations~\eqref{eqVarCoefTelegraphEq} are used to model a wide variety of phenomena in physics,
chemistry, mathematical biology etc. For the case $f (x) = 1$ and
$K(u)=0$ equation~\eqref{eqVarCoefTelegraphEq} can be used to describe the flow of
one-dimensional gas, longitudinal wave propagation on a moving
threadline and dynamics of a finite nonlinear string and so on~\cite{Ames1965/72,Ames1981}.
When $K(u) = 0$ this equation describes the longitudinal
vibrations of an elastic and non-homogeneous taut string or bar~\cite{Torrisi&Valenti1985}.
The outstanding representative of the class of equations~\eqref{eqVarCoefTelegraphEq}
is the nonlinear telegraph equation that is the mathematical model
for a large number of physical phenomena. (For more details refer
to~\cite{Ames1965/72,Malfliet1992}.)

Historically, there are a number of papers contributed to the
studies of Lie groups of transformations of various class of
$(1+1)$-dimensional nonlinear wave equations and their individual
members. Probably, Barone {\it et al}~\cite{Barone&Esposito&Magee&Scott1971} was the first study of the
following nonlinear wave equation
\[
u_{tt}=u_{xx}+F(u),
\]
by means of symmetry method, this equation was also studied by
Kumei~\cite{Kumei1975} and Pucci {\it  et al}~\cite{Pucci&Salvatori1986} subsequently. Motivated by a
number of physical problems, Ames {\it  et al}~\cite{Ames1965/72,Ames1981} investigated
group properties of quasi-linear hyperbolic equations of the form
\begin{equation}\label{eqWaveEq1}
u_{tt}=[f(u)u_x]_x.
\end{equation}
Later, their investigation was generalized in~\cite{Torrisi&Valenti1985,Donato1987,Ibragimov&Torrisi&Valenti1991} to
equations of the following forms respectively
\begin{gather*}
u_{tt}=[f(x, u)u_x]_x,\quad
u_{tt}=[f(u)u_x+g(x,u)]_x,\quad\mbox{and}\quad
u_{tt}=f(x, u_x)u_{xx}+g(x,u_x).
\end{gather*}
The alternative form of equation~\eqref{eqWaveEq1} was also investigated by Oron and Rosenau~\cite{Oron&Rosenau1986}
and Suhubi and Bakkaloglu~\cite{Suhubi&Bakkaloglu1991}.
Arrigo~\cite{Arrigo1991} classified the equations
\[
u_{tt}=u_x^{m}u_{xx}+F(u).
\]
Furthermore, classification results for the equation
\[
u_{tt}+K(u)u_t=[F(u)u_x]_x
\]
can be found in~\cite{Ibragimov1994V1}. An expand
form of the latter equation
\[
u_{tt}+K(u)u_t=[F(u)u_x]_x+H(u)u_x
\]
was studied by Kingston and Sophocleous~\cite{Kingston&Sophocleous2001}.
Recently, Lahno, Zhdanov and Magda~\cite{Lahno&Zhdanov&Magda2006} presented
the most extensive list of symmetries of the equations
\[
u_{tt}=u_{xx}+F(t, x, u, u_x)
\]
by using the infinitesimal Lie method, the technique of
equivalence transformations and the theory of classification of
abstract low-dimensional Lie algebras.
There are also some papers~\cite{Oron&Rosenau1986,Chikwendu1981,Gandarias&Torrisi&Valenti2004,Pucci1987}
devoted to the group classification of the equation of
the following form
\begin{gather*}
u_{tt}=F(u_{xx}),
\quad
u_{tt}=F(u_x)u_{xx}+H(u_x),\quad\mbox{and}\quad u_{tt}+\lambda
u_{xx}=g(u,u_x).
\end{gather*}
It worthwhile mentioned that the equations
\[
u_{tt}=(F(u)u_x)_{x}+H(u)u_x
\]
together with its equivalent potential systems have also been studied by
Bluman {\it et al}~\cite{Bluman&Temuerchaolu&Sahadevan2005,Bluman&Temuerchaolu2005a,
Bluman&Temuerchaolu2005b,Bluman&Cheviakov&Ivanova2005}.
In their a series of papers, many interesting results (especially for case of power nonlinearities)
including Lie point and nonlocal symmetries classification and conservation laws of the four
equivalent systems were systematically investigated.

From the above introduction, it is easy to see that equation~\eqref{eqVarCoefTelegraphEq}
is different from any aforementioned ones. However, equation~\eqref{eqVarCoefTelegraphEq}
is a generalization of many physically important systems, thus
there is essential interest in investigating them from a unified and
group theoretical point. The ultimate goal of this paper is to
present an extended group analysis and to find additional
equivalence transformations and exact solutions of equations~\eqref{eqVarCoefTelegraphEq}.
A lot of new interesting cases of extensions of the maximal Lie
symmetry group and cases with high-dimensional spaces of conservation laws
were obtained for these equations.

Problems of general group classification, except for really
trivial cases, are very difficult. This can be illustrated by the
multitude of papers where such a general classification problem is
solved incorrectly or incompletely. There are also many papers on
``preliminary group classification'' where authors list some cases
with new symmetry but do not claim that the general classification
problem is solved completely. For this reason, finding an
effective approach to simplification is essentially equivalent to
showing the feasibility of solving the problem at all. Recently,
based on the investigation of the specific compatibility of
classifying conditions, Nikitin and Popovych~\cite{Nikitin&Popovych2001} developed an
effective tool (we refer it as compatibility method) for solving
the group classification problem of nonlinear Schr\"{o}dinger
equation. Their method has been applied to investigating a number
of different group classification problem~\cite{Nikitin&Popovych2001,Popovych&Cherniha2001,
Boyko&Popovych2001,Popovych&Ivanova2004NVCDCEs,Ivanova&Sophocleous2006,Vasilenko&Yehorchenko2001}. In particular,
in~\cite{Popovych&Ivanova2004NVCDCEs} Popovych and Ivanova extended the method to complete group
classification of nonlinear diffusion-convection equations by
further considering the so called additional equivalence
transformations. However, to the best our knowledge, there are no
any result about application the compatibility method to equation~\eqref{eqVarCoefTelegraphEq}.
Therefore, the paper is one of new application of the
compatibility method to the problem of group classification.
Although the authors' debt to the works of Nikitin, Popovych and
Ivanova~\cite{Nikitin&Popovych2001,Popovych&Ivanova2004NVCDCEs} is evident, the results of group classification
of class~\eqref{eqVarCoefTelegraphEq} presented in this work seem to be new. Hence, these
will lead to some explicit applications in Physics and Engineer.

The rest of this paper is organized as follows.
Since the case $f(x)=1$ has a great variety of applications
and has been investigated earlier by
a number of authors, we collect results for this class together in
Section~\ref{SectionOnGrClasConstCoef}.
In Section~\ref{SectionOnGrClasRes} we
present the results of the complete group classification of
class~\eqref{eqVarCoefTelegraphEq}. Some additional
equivalence transformations are considered in
Section~\ref{SectionOnAdditEquivTr}, where we also present
the result of group classification of class~\eqref{eqVarCoefTelegraphEq}
with respect to the set of point transformations.
The result of the group classification is used to find exact
solutions of equations from class~\eqref{eqVarCoefTelegraphEq}
(Section~\ref{SectionOnExactSol}).
Ibid we construct functionally separation solutions for some equations~\eqref{eqVarCoefTelegraphEq}.
A natural continuation of group analysis of equations~\eqref{eqVarCoefTelegraphEq} is investigation of
their conservation laws. More precisely, in Section~\ref{SectionOnConsLaws} we construct
the local conservation laws of equations of form~\eqref{eqVarCoefTelegraphEq} having characteristics
of order~0. Finally, some conclusion and
discussion are given in Section~\ref{SectionOnConclusion}.

\section{Group classification for
the subclass with \mathversion{bold}$f(x)=1$\mathversion{normal}}\label{SectionOnGrClasConstCoef}

Class~\eqref{eqVarCoefTelegraphEq} includes a subclass of equations of the general form
\begin{equation}\label{eqConstCoefTelegraphEq}
u_{tt} = (H(u)u_x)_x + K(u)u_x
\end{equation}
(i.e., the function $f$ is assumed to be equal to $1$ identically).
These equations (called the {\em nonlinear telegraph equations}) are important for applications.
Symmetry properties of class~\eqref{eqConstCoefTelegraphEq}
were studied by a number of authors
(see~\cite{Bluman&Temuerchaolu&Sahadevan2005,Ibragimov1994V1,Kingston&Sophocleous2001,Gandarias&Torrisi&Valenti2004}
for details).
However, in all the above references the results of group classification of class~\eqref{eqConstCoefTelegraphEq}
are presented in implicit form only.
Therefore we single out the results of
the group classification of equations~\eqref{eqConstCoefTelegraphEq} from classification
of class~\eqref{eqVarCoefTelegraphEq}.


\begin{theorem}
The Lie algebra of the kernel of principal
groups of~\eqref{eqConstCoefTelegraphEq} is $A_1^{\ker}=\langle\partial_t,\, \partial_x\rangle $.
\end{theorem}

\begin{theorem}
The Lie algebra of the equivalence group~$G_1^{\sim}$ for class~\eqref{eqConstCoefTelegraphEq} is
\[
A_1^{\sim}=\langle\partial_t,\,\partial_x,\,
 \partial_u,\, u\partial_u,\,t\partial_t-2H\partial_H-2K\partial_K,\,
 x\partial_x +2H\partial_H+K\partial_K\rangle .
\]
\end{theorem}

Any transformation from $G_1^{\sim}$ for the class~\eqref{eqConstCoefTelegraphEq} is
\[
\tilde t = t \epsilon_4+\epsilon_1,\quad
\tilde x = x \epsilon_5+\epsilon_2,\quad
\tilde u = u \epsilon_6 + \epsilon_3,\quad
\tilde H=H \epsilon_4^{-2}\epsilon_5^{2},\quad
\tilde K= K \epsilon_4^{-2}\epsilon_5,
\]
where $\epsilon_1, \ldots , \epsilon_7$ are arbitrary constants,
$\epsilon_4\epsilon_5\epsilon_6\neq 0$.

\begin{theorem}\label{TheorOnGrClasResF1}
The complete set of
$G_1^{\sim}$-inequivalent extensions of $A^{\max} \neq A^{\ker}$
for equation~\eqref{eqConstCoefTelegraphEq} is exhausted by ones given in table~\ref{TableGrClasF1}.
\end{theorem}

\setcounter{tbn}{0}

\begin{center}\footnotesize\renewcommand{\arraystretch}{1.15}
Table~\refstepcounter{table}\label{TableGrClasF1}\thetable. Case of $f(x)=1$ \\[1ex]
\begin{tabular}{|l|c|c|l|}
\hline
N & $H(u)$ & $K(u)$  & \hfil Basis of A$^{\max}$ \\
\hline
\refstepcounter{tbn}\label{CaseGrClasF1ForallHForAllK}\thetbn & $\forall$ & $\forall$ & $\partial_t,\, \partial_x$ \\
\refstepcounter{tbn}\label{CaseF1ForAllHK0}\thetbn &
$\forall$ & $0$
& $\partial_t,\,  \partial_x,\, t\partial_t+x\partial_x$  \\
\refstepcounter{tbn}\label{CaseF1HexpKexp}\thetbn & $e^{\mu u}$ & $e^{\nu u}$ &
 $\partial_t,\, \partial_x,\, (\mu-2\nu)t\partial_t+2(\mu-\nu)x\partial_x+2\partial_u$  \\
\refstepcounter{tbn}\label{CaseF1HexpK1}\thetbn &
$e^{u}$ & $1$ & $\partial_t,\,\p_x,\,  t\partial_t+2x\partial_x+2\partial_u$  \\
\refstepcounter{tbn}\label{CaseF1HexpK0}\thetbn & $e^{u}$ & $0$ &
$\partial_t,\, \partial_x,\, t\partial_t-2\partial_u,\, x\partial_x+2\p_u$  \\
\refstepcounter{tbn}\label{CaseF1HpowerKpower}\thetbn & $|u|^{\mu}$ & $|u|^{\nu }$ &
$\partial_t,\, \partial_x,\, (\mu-2\nu)t\partial_t+2(\mu-\nu)x\partial_x+2u\partial_u$  \\
\refstepcounter{tbn}\label{CaseF1HpowerK1}\thetbn & $|u|^{\mu}$ & $1$ &
$\partial_t,\, \partial_x,\, \mu t\partial_t+2\mu x\partial_x+2u\partial_u$  \\
\refstepcounter{tbn}\label{CaseF1HpowerK0}\thetbn a& $|u|^{\mu}$ & $0$  &
$\partial_t,\, \partial_x,\, \mu t\partial_t-2u\partial_u,\, \mu x\partial_x+2u\partial_u$\\
\thetbn b& $u^{-2}$ & $u^{-2}$  &
$\partial_t,\, \partial_x,\, t\partial_t+u\partial_u,\, e^{-x}(\partial_x+u\partial_u)$\\
\refstepcounter{tbn}\label{CaseF1Hu-4Ku-4}\thetbn & $u^{-4}$ & $u^{-4}$ &
$\partial_t,\, \partial_x,\, 2t\partial_t+u\partial_u,\, t^2\partial_t+tu\partial_u$  \\
\refstepcounter{tbn}\label{CaseF1Hu-4K0}\thetbn & $u^{-4}$ & $0$ &
$\partial_t,\, \partial_x,\, 2t\partial_t+u\partial_u,\, 2x\partial_x-u\partial_u,\, t^2\partial_t+tu\partial_u$  \\
\refstepcounter{tbn}\label{CaseF1Hu-43K0}\thetbn &$u^{-4/3}$ & $0$ &
$\partial_t,\, \partial_x,\, 2t\partial_t+3u\partial_u,\, 2x\partial_x-3u\partial_u,\, x^2\partial_x-3xu\partial_u$  \\
\hline
\end{tabular}
\end{center}
{\footnotesize
 Here $(\mu,\nu)\ne(0,0)$, $(\mu,\nu)\in\{(1,0),(0,1)\}\!\!\mod G_1^{\sim}$ in case~\ref{CaseF1HexpKexp},
 $\mu\ne0,-4,-4/3$ in case~\ref{CaseF1HpowerK0}.
Case~\ref{CaseF1HpowerK0}b is reduced to case~\ref{CaseF1HpowerK0}a by means of transformation $\tilde t=t,$ $\tilde x=e^x,$ $\tilde u=e^{-x}u.$
 }

\begin{remark}
The most similar form of classification results have been
presented in~\cite{Bluman&Temuerchaolu&Sahadevan2005}. Note that
the point symmetries of some special classes of
equation~\eqref{eqConstCoefTelegraphEq} adduced in table~III
of~\cite{Bluman&Temuerchaolu&Sahadevan2005} can be obtained,
respectively, by appropriately scaling $t$ and $u$ or taking some
appropriately values for the parameters $\mu$ and $\nu$ in cases
\ref{TableGrClasF1}.\ref{CaseF1HexpKexp} and
\ref{TableGrClasF1}.\ref{CaseF1HpowerKpower}. For example, the
classes A and E of table~III
in~\cite{Bluman&Temuerchaolu&Sahadevan2005} can be obtained by
setting $\mu=\alpha+1$, $\nu=1$ and $\mu=\alpha$, $\nu=1$
respectively in case \ref{TableGrClasF1}.\ref{CaseF1HexpKexp}.
\end{remark}

\begin{remark}
The proof of theorem~\ref{TheorOnGrClasResF1} follows directly from the results of the next section.
\end{remark}

\section{Results of Classification}\label{SectionOnGrClasRes}

Following the above-mentioned algorithm we are looking for an
infinitesimal operator in the form
\begin{equation}\label{OperatorLieSym}
Q=\tau(t, x, u)\partial_t+\xi(t, x, u)\partial_x+\eta(t, x,u)\partial_u
\end{equation}
which corresponds to a one-parameter Lie
group of local transformations and keeps the equation~\eqref{eqVarCoefTelegraphEq} invariant.
The classical infinitesimal Lie invariance criterion for equation~\eqref{eqVarCoefTelegraphEq}
to be invariant with respect to the operator~\eqref{OperatorLieSym} read as
\begin{equation}\label{sysDetEqTelEqGenForm}
\pr^{(2)}Q(\triangle)\mid_{\triangle=0}=0,\qquad
 \triangle=f(x)u_{tt}-(H(u)u_x)_x-K(u)u_x.
\end{equation}
Here $\pr^{(2)}Q$ is the usual second order prolongation~\cite{Olver1986,Ovsiannikov1982} of
the operator~\eqref{OperatorLieSym}.
Substituting the coefficients of~$\pr^{(2)}Q$ into~\eqref{sysDetEqTelEqGenForm} yields the
following determining equations for $\tau$, $\xi$ and $\eta$:
\begin{equation}\label{sysDetEqTelEqComp}
\begin{array}{ll}
\tau_x=\tau_u=\xi_t=\xi_u=\eta_{uu}=0,\quad
H\eta_{xx}+K\eta_x-f\eta_{tt}=0,\\
\frac{f_x}{f}H\xi-2\tau_tH-\eta
H_u+2H\xi_x=0,\\
H\xi_{xx}-2H_u\eta_x-\eta K_u-2\tau_t
K+\frac{f_x}{f}K\xi+\xi_xK-2\eta_{xu}H=0,\\
2\eta_{tu}f-\tau_{tt}f=0,\quad
2H\xi_{xu}-2\tau_tH_u+\frac{f_x}{f}H_u\xi+2H_u\xi_x-\eta_uH_u-\eta
H_{uu}=0.
\end{array}
\end{equation}
Investigating the compatibility of system~\eqref{sysDetEqTelEqComp} we find that the
final equation of system~\eqref{sysDetEqTelEqComp} is an identity (substituting the
second equation of system~\eqref{sysDetEqTelEqComp} to the final one can yield this
conclusion). With this condition, system~\eqref{sysDetEqTelEqComp} can be rewritten in
the form
\begin{gather}
\tau_x=\tau_u=\xi_t=\xi_u=\eta_{uu}=0,\quad
2\eta_{tu}=\tau_{tt},\label{sysDetEqTelEq_1}\\
2(\xi_x-\tau_t)+\frac{f_x}{f}\xi=\frac{H_u}{H}\eta,\label{sysDetEqTelEq_2}\\
H\eta_{xx}+K\eta_x-f\eta_{tt}=0,\label{sysDetEqTelEq_3}\\
(H_uK-K_uH)\frac{\eta}{H}-K\xi_x-2H_u\eta_x+H(\xi_{xx}-2\eta_{xu})=0.\label{sysDetEqTelEq_4}
\end{gather}
Equations~\eqref{sysDetEqTelEq_1} do not contain arbitrary elements. Integration of
them yields
\begin{equation}\label{CoefsSymIntegr}
\tau =\tau(t),\quad \xi=\xi(x),\quad \eta= \eta^1(t, x)u + \eta^0(t, x),
\quad \eta^1(t, x)=\frac{1}{2}\tau_t+\alpha(x).
\end{equation}

Thus, group classification of~\eqref{eqVarCoefTelegraphEq} reduces to solving classifying
conditions~\eqref{sysDetEqTelEq_2}--\eqref{sysDetEqTelEq_4}.

Splitting system~\eqref{sysDetEqTelEq_2}--\eqref{sysDetEqTelEq_4} with respect to the arbitrary elements
and their non-vanishing derivatives gives the equations
$\tau_t=0$, $\xi = 0$, $\eta= 0$
for the coefficients of the operators from~$A^{\ker}$ of~\eqref{eqVarCoefTelegraphEq}.
As a result, the following theorem is true.

\begin{theorem}\label{TheorOnKernel}
The Lie algebra of the kernel of principal groups of~\eqref{eqVarCoefTelegraphEq} is
$A^{\ker}=\langle\partial_t\rangle$.
\end{theorem}

The next step of the algorithm of group classification is finding
equivalence transformations of class~\eqref{eqVarCoefTelegraphEq}.
An equivalence transformation is a nondegenerate change of the variables $t$, $x$ and~$u$
taking any equation of the form~\eqref{eqVarCoefTelegraphEq} into an equation of the
same form, generally speaking, with different $f(x)$, $H(u)$ and~$K(u)$.
The set of all equivalence transformations forms the
equivalence group $G^{\sim}$. To find the connected component of the unity of~$G^{\sim}$,
we have to investigate Lie symmetries of the system that consists of
equation~\eqref{eqVarCoefTelegraphEq} and some additional conditions, that is to say we
must seek for an operator of the Lie algebra~$A^{\sim}$ of~$G^{\sim}$ in the form
\begin{equation}\label{OperatorEquivTr}
X=\tau \partial_t+\xi \partial_x+\eta \partial_u+\pi
\partial_f+\rho \partial_H+ \varphi \partial_K
\end{equation}
from the invariance criterion of~\eqref{eqVarCoefTelegraphEq} applied to the system:
\begin{equation}\label{sysForEquivTr}
\begin{array}{l}
f(x)u_{tt}=(H(u)u_x)_x+K(u)u_x,\\[1ex]
f_t=f_u=0,\quad H_t=H_x=0,\quad K_t=K_x=0.
\end{array}
\end{equation}
Here $u$, $f$, $H$ and $K$ are
considered as differential variables: $u$ on the space $(t,x)$ and
$f$, $H$, $G$ on the extended space $(t, x, u)$. The coordinates
$\tau$, $\xi$, $\eta$ of the operator~\eqref{OperatorEquivTr} are sought as functions of
$t$, $x$, $u$ while the coordinates $\pi$, $\rho$, $\varphi$ are sought as
functions of $t$, $x$, $u$, $f$, $H$, $K$.

The invariance criterion of system~\eqref{sysForEquivTr} yields
the following determining equations for $\tau$, $\xi$, $\eta$, $\pi$,
$\rho$  and $\varphi$:
\begin{equation}\label{sysDetEqEquivTr}
\begin{array}{ll}
\tau_x=\tau_u=\xi_t=\xi_u=\eta_x=\eta_u=0, \quad\quad
\tau_{tt}=\eta_{uu}=0,\\
\pi_t=\pi_u=\pi_H=\pi_K,\quad\quad
\rho_t=\rho_x=\rho_u=\rho_f=\rho_K,\quad\quad\varphi_t=\varphi_x=\varphi_f=0,\\
\frac{\pi}{f}+2\xi_x-2\tau_t-\rho_{H}=0,\\
\frac{\pi}{f}+2\xi_x-2\tau_t=\frac{\rho}{H},\\
\xi_{xx}H+(\frac{\pi}{f}+\xi_x-2\tau_t)K-\varphi=0.
\end{array}
\end{equation}
After easy calculations we find from~\eqref{sysDetEqEquivTr}
\[
\begin{array}{ll}\tau=c_1+c_4t, \quad\quad \xi=c_2+c_5x,\quad\quad
\eta=c_3+c_6u,\\
\pi=(c_7+2c_4-2c_5)f,\quad\quad \rho=c_7H,
\quad\quad\varphi=(c_7-c_5)K,
\end{array}
\]
where $c_1,\ldots , c_7$ are arbitrary constants. Thus, we obtain
the following statement.

\begin{theorem}\label{TheorOnEquivAlg}
The Lie algebra of $G^{\sim}$ for class~\eqref{eqVarCoefTelegraphEq}
is
\[
A^{\sim}=\langle\partial_t, \partial_x, \partial_u,
t\partial_t+2f\partial_f, x\partial_x-2f\partial_f-K\partial_K,
u\partial_u, f\partial_f+H\partial_u+K\partial_K\rangle .
\]
\end{theorem}

Continuous equivalence transformations of class~\eqref{eqVarCoefTelegraphEq} are generated by the
operators from~$A^{\sim}$. For class~\eqref{eqVarCoefTelegraphEq} there also exists a non-trivial
group of discrete equivalence transformations generated by four
involutive transformations of alternating sign in the sets $\{t\},
\{x,K\},$ $ \{u\}$ and $\{f,H,K\}$.
Therefore, $G^{\sim}$ contains the following continuous
transformations:
\[
\begin{array}{ll}
\tilde t = t {\epsilon_4}+\epsilon_1,\quad \tilde x = x {\epsilon_5}+\epsilon_2,\quad
\tilde u = u {\epsilon_6} + \epsilon_3,\quad
\tilde f = f {\epsilon_4^2\epsilon_5^{-2}\epsilon_7},\quad
\tilde H=H {\epsilon_7},\quad \tilde K= K{\epsilon_7\epsilon_5^{-1}},
\end{array}
\]
where $\epsilon_1, \ldots,\epsilon_7$ are arbitrary constants.

\begin{theorem}\label{TheorOnGrClasRes}
A complete set of inequivalent equations~\eqref{eqVarCoefTelegraphEq}
with respect to the transformations from $G^{\sim}$ with $A^{\max}
\neq A^{\ker}$ is exhausted by cases given in tables~\ref{TableGrClasForAllH}--\ref{TableGrClasHpower}.
\end{theorem}

\setcounter{tbn}{0}

\begin{center}\footnotesize\renewcommand{\arraystretch}{1.15}
Table~\refstepcounter{table}\label{TableGrClasForAllH}\thetable. Case of $\forall H(u)$ \\[1ex]
\begin{tabular}{|l|c|c|c|l|}
\hline
N & $K(u)$ & $f(x)$  & \hfil Basis of A$^{\max}$ \\
\hline
\refstepcounter{tbn}\label{CaseForAllHForAllKForallF}\thetbn & $\forall$ & $\forall$ & $\p_t$ \\
\refstepcounter{tbn}\label{CaseForAllHForAllKFexp}\thetbn a
&$\forall$ & $e^{\epsilon x}$ &$\p_t,\, \epsilon t\partial_t+2\partial_x$ \\
\thetbn b & $H$ & $e^{-2x-\gamma e^{-x}}$  & $\partial_t,\,  \gamma t\partial_t+2e^x\partial_x$ \\
\thetbn c & $H$ & $e^{-2x}(e^{-x}+\gamma)^{\lambda}$ & $\partial_t,\,(\lambda+2)t\partial_t-2(1+\gamma e^x)\partial_x$\\
\thetbn d & 0 & $|x|^{\lambda}$  & $\partial_t,\,  (\lambda+2)t\partial_t+2x\partial_x$ \\
\refstepcounter{tbn}\label{CaseForAllHK0F1}\thetbn a& 0 & 1  & $\partial_t,\,   \partial_x,\, t\partial_t+x\partial_x$\\
\thetbn b& $H$ & $e^{-2 x}$  & $\partial_t,\,  t\partial_t-\partial_x,\,  e^{x}\partial_x$ \\
\hline
\end{tabular}
\end{center}
{\footnotesize
Here $\gamma,\, \lambda\neq 0$, $\epsilon\in\{0, 1\}\!\! \mod G^{\sim}$,
$\gamma=\pm 1\!\! \mod G^{\sim}$.\\
Additional equivalence transformations:\\ \setcounter{casetran}{0}
\refstepcounter{casetran}\thecasetran.
\ref{CaseForAllHForAllKFexp}b $\to$ \ref{CaseForAllHForAllKFexp}a$(K=0, \epsilon=1)$:\quad
$\tilde t=t$, $\tilde x=-\gamma e^{-x}$, $\tilde u=u;$\\
\refstepcounter{casetran}\thecasetran.
\ref{CaseForAllHForAllKFexp}c($\lambda \neq -2$) $\to$
\ref{CaseForAllHForAllKFexp}a$(K=-H/(\lambda+2),\
\epsilon=1)$:\quad
$\tilde t=t$, $\tilde x=(\lambda+2)\ln|\gamma+e^{-x}|$, $\tilde u=u;$\\
\phantom{2.}{} \ref{CaseForAllHForAllKFexp}c($\lambda= -2$) $\to$
\ref{CaseForAllHForAllKFexp}a$(K=-H, \epsilon=0)$:\quad
$\tilde t=t$, $\tilde x=\ln|\gamma+e^{-x}|$, $\tilde u=u;$\\
\refstepcounter{casetran}\thecasetran.
\ref{CaseForAllHForAllKFexp}d($\lambda\neq -2)$ $\to$
\ref{CaseForAllHForAllKFexp}a$(K=-H/(\lambda+2),\
\epsilon=1)$:\quad
$\tilde t=t$, $\tilde x=(\lambda+2)\ln|x|$, $\tilde u=u;$\\
\phantom{2.}{} \ref{CaseForAllHForAllKFexp}d($\lambda= -2$) $\to$
\ref{CaseForAllHForAllKFexp}a$(K=-H,\ \epsilon=0)$:\quad
$\tilde t=t$, $\tilde x=\ln|x|$, $\tilde u=u;$\\
\refstepcounter{casetran}\thecasetran. \ref{CaseForAllHK0F1}b $\to$ \ref{CaseForAllHK0F1}a:\quad
$\tilde t=t$, $\tilde x=e^{-x}$, $\tilde u=u$. }

\setcounter{tbn}{0}

\begin{center}\footnotesize\renewcommand{\arraystretch}{1.15}
Table~\refstepcounter{table}\label{TableGrClasHexp}\thetable. Case of $H(u)=e^{\mu u}$ \\[1ex]
\begin{tabular}{|l|c|c|c|l|}
\hline
N & $\mu$ & $K(u)$ & $f(x)$  & \hfil Basis of A$^{\max}$ \\
\hline
\refstepcounter{tbn}\label{CaseHexpKexpnuuFxlambda}\thetbn &
$\forall$ &  $e^{\nu u}$ & $|x|^\lambda$ &
$\p_t,\, [\lambda(\mu-\nu)+(\mu-2\nu)]t\partial_t+2(\mu-\nu)x\partial_x+2\partial_u $ \\
\refstepcounter{tbn}\label{CaseHexpKexpnuuF1}\thetbn &
$\forall$ & $e^{\nu u}$ &
$1$ & $\partial_t,\, \partial_x,\, (\mu-2\nu)t\partial_t+2(\mu-\nu)x\partial_x+2\partial_u$  \\
\refstepcounter{tbn}\label{CaseHexpK1Fxlambda}\thetbn &
$1$ & $1$ &
$|x|^\lambda$ & $\partial_t,\,  (\lambda+1)t\partial_t+2x\partial_x+2\partial_u$  \\
\refstepcounter{tbn}\label{CaseHexpK1F1}\thetbn &
$1$ & $1$ &
$1$ & $\partial_t,\,\p_x,\,  t\partial_t+2x\partial_x+2\partial_u$  \\
\refstepcounter{tbn}\label{CaseHexpK0expuForAllF}\thetbn &
$1$ & $\epsilon e^{u}$ &
$\forall$ & $\partial_t,\, t\partial_t-2\partial_u$  \\
\refstepcounter{tbn}\label{CaseHexpK0Ff1}\thetbn a &
$1$ & $0$ & $f^{1}(x)$ &
$\partial_t,\,t\partial_t-2\partial_u,\,
\alpha t\partial_t+2(\beta x^2+\gamma_1x+\gamma_0)\partial_x+2\beta x\partial_u$  \\
\thetbn b & $1$ & $e^{u}$ & $f^{2}(x)$ &
$\partial_t,\,t\partial_t-2\partial_u,\,
\alpha t\partial_t-2(\gamma_0e^x+\gamma_1+\beta e^{-x})\partial_x+2\beta e^{-x}\partial_u$  \\
\refstepcounter{tbn}\label{CaseHexpK0F1}\thetbn a &
$1$ & $0$ & $1$ &
$\partial_t,\, t\partial_t-2\partial_u,\, x\partial_x+2\p_u,\, \partial_x$  \\
\thetbn b & $1$ & $e^{u}$ & $e^{-2x}$ &
$\partial_t,\, t\partial_t-2\partial_u,\, \partial_x-2\partial_u,\, e^{x}\partial_x$  \\
\thetbn c & $1$ & $0$ & $x^{-3}$ &
$\partial_t,\,t\partial_t-2\partial_u,\,  x\partial_x-\partial_u,\, x^2\partial_x+x\partial_u$  \\
\thetbn d & $1$ & $e^{u}$ & $e^{-2x}(e^{-x}+\gamma)^{-3}$ &
$\partial_t,\, t\partial_t-2\partial_u,\, t\p_t+2(1+\gamma e^x)\partial_x,\,
(e^{-x}+\gamma)^{2}e^x\partial_x-(e^{-x}+\gamma)\partial_u$  \\
\hline
\end{tabular}
\end{center}
{\footnotesize
Here $(\mu,\nu)\ne(0,0)$, $(\mu,\nu)\in\{(1,0),(0,1)\}\!\!\mod G^{\sim}$;
$\gamma=\pm 1\!\! \mod G^{\sim}$;
$\lambda \neq 0$; $\epsilon\in \{0, 1\}\!\!\mod G^{\sim}$;
$\alpha$, $\beta$, $\gamma_1$, $\gamma_0=\const$ and
\[
f^1(x)=\exp\left\{\int \frac{-3\beta x-2\gamma_1+\alpha}{\beta
x^2+\gamma_1 x+\gamma_0} dx\right\},\qquad
f^2(x)=\exp\left\{\int \frac{\beta
e^{-x}-\alpha-2\gamma_0e^x}{\beta
e^{-x}+\gamma_1+\gamma_0} dx\right\}.
\]
Additional equivalence transformations:\\ \setcounter{casetran}{0}
\refstepcounter{casetran}\thecasetran. \ref{CaseHexpK0Ff1}b $\to$ \ref{CaseHexpK0Ff1}a:\quad
$\tilde t=t$, $\tilde x=e^{-x}$, $\tilde u=u;$\\
\refstepcounter{casetran}\thecasetran. \ref{CaseHexpK0F1}b $\to$ \ref{CaseHexpK0F1}a:\quad
$\tilde t=t$, $\tilde x=e^{-x}$, $\tilde u=u;$\\
\refstepcounter{casetran}\thecasetran. \ref{CaseHexpK0F1}c $\to$ \ref{CaseHexpK0F1}a\quad
$\tilde t=t\sign x$, $\tilde x=1/x$, $\tilde u=u-\ln|x|;$\\
\refstepcounter{casetran}\thecasetran. \ref{CaseHexpK0F1}d $\to$ \ref{CaseHexpK0F1}a:\quad
$\tilde t=t\sign(\gamma+e^{-x})$, $\tilde x=1/(\gamma+e^{-x})$, $\tilde u=u-\ln|(\gamma+e^{-x})|.$
}

\setcounter{tbn}{0}

\begin{center}\footnotesize\renewcommand{\arraystretch}{1.15}
Table~\refstepcounter{table}\label{TableGrClasHpower}\thetable. Case of $H(u)=|u|^{\mu}$ \\[1ex]
\begin{tabular}{|l|c|c|c|l|}
\hline
N & $\mu$ & $K(u)$ & $f(x)$  & \hfil Basis of A$^{\max}$ \\
\hline
\refstepcounter{tbn}\label{CaseHpowerKpowerFpower}\thetbn &
$\forall$ & $|u|^{\nu}$ &
$|x|^\lambda$ & $\partial_t,\, [\lambda(\mu-\nu)+(\mu-2\nu)]t\partial_t+2(\mu-\nu)x\partial_x+2u\partial_u$  \\
\refstepcounter{tbn}\label{CaseHpowerKpowerF1}\thetbn &
$\forall$ & $|u|^{\nu }$ & $1$ & $\partial_t,\,
\partial_x,\, (\mu-2\nu)t\partial_t+2(\mu-\nu)x\partial_x+2u\partial_u$\\
\refstepcounter{tbn}\label{CaseHpowerK1Fpower} \thetbn &
$\forall$ & $1$ & $|x|^\lambda$ &
$\partial_t,\, (\lambda+1)\mu t\partial_t+2\mu x\partial_x+2u\partial_u$\\
\refstepcounter{tbn}\label{CaseHpowerK1F1} \thetbn &
$\forall$ & $1$ & $1$ & $\partial_t,\, \partial_x,\,
\mu t\partial_t+2\mu x\partial_x+2u\partial_u$\\
\refstepcounter{tbn}\label{CaseHpowerK0HForAllF}\thetbn &
 $\neq -4$ & $ |u|^{\mu}$ & $\forall$
 & $\partial_t,\, \mu t\partial_t-2u\partial_u$  \\
\refstepcounter{tbn}\label{CaseHpowerK0Ff3}\thetbn a &
$ \neq -4$ & $0$ & $f^{3}(x)$ &
$\partial_t,\, \mu t\partial_t-2u\partial_u,\,
\alpha t\partial_t + 2[(\mu+1)\beta x^2+\gamma_1 x+\gamma_0]\partial_x+2\beta xu\partial_u$  \\
\thetbn b & $\neq -4$ & $|u|^{\mu}$ & $f^{4}(x)$ &
$\partial_t,\, \mu t\partial_t-2u\partial_u,$\\
&  & & & $ \alpha t\partial_t -2[(\mu+1)\beta e^{-x}+\gamma_1+\gamma_0 e^{x}]\partial_x+2\beta e^{-x}u\partial_u$ \\
\refstepcounter{tbn}\label{CaseHpowerK0F1}\thetbn a &
$\neq -4, -\frac{4}{3}$ & $0$ & $1$ &
$\partial_t,\, \mu t\partial_t-2u\partial_u,\,\partial_x,\, \mu x\partial_x+2u\partial_u$\\
\thetbn b & $\neq -4, -\frac{4}{3}$ & $|u|^{\mu}$ & $e^{-2x}$ &
$\partial_t,\, \mu t\partial_t-2u\partial_u,\,\mu \partial_x-2u\partial_u,\, e^x\partial_x$\\
\thetbn c & $\neq -4, -\frac{4}{3}, -1$ & $0$ &
$|x|^{-\frac{3\mu+4}{\mu+1}}$ &
$\partial_t,\, \mu t\partial_t-2u\partial_u,\,
\mu(\mu+1)x\partial_x-(\mu+2)u\partial_u,$\\
&  & & & $(\mu+1)x^2\partial_x+xu\partial_u$  \\
\thetbn d & $\neq -4, -\frac{4}{3}, -1$ & $|u|^{\mu}$ &
$\frac{e^{-2x}}{(e^{-x}+\gamma)^{\frac{3\mu+4}{\mu+1}}}$ &
$\partial_t,\, \mu t\partial_t-2u\partial_u,\,
(\mu+2)t\partial_t +2(\mu+1)(e^{-x}+\gamma)e^{x}\partial_x,$\\
&  & & & $(\mu+1)(e^{-x}+\gamma)^2e^x\partial_x-(e^{-x}+\gamma)u\partial_u$  \\
\thetbn e & $-1$ & $0$ & $e^{\gamma x}$ &
$\partial_t,\,t\partial_t+2u\partial_u,\,\partial_x-\gamma u\partial_u,\,
 t\partial_t+x\partial_x-\gamma xu\partial_u$   \\
\thetbn f & $-1$ & $u^{-1}$ & $e^{-2x+\gamma e^{-x}}$ &
$\partial_t,\, t\partial_t+2u\partial_u,\,
e^x\partial_x-\gamma u\partial_u,\, t\partial_t-\partial_x+\gamma e^{-x}u\partial_u$   \\
\thetbn g& $-2$ & $u^{-2}$ & $1$  &
$\partial_t,\, \partial_x,\, t\partial_t+u\partial_u,\, e^{-x}(\partial_x+u\partial_u)$\\
\refstepcounter{tbn}\label{CaseHu-4K0HForallF} \thetbn &
$-4$ & $ u^{-4}$ & $\forall$ & $\partial_t,\, 2t\partial_t+u\partial_u,\, t^2\partial_t+tu\partial_u$  \\
\refstepcounter{tbn}\label{CaseHu-4K0Ff3} \thetbn a
& $-4$ & $0$ & $f^3(x)|_{\mu=-4}$ & $\partial_t,\,
2t\partial_t+u\partial_u,\, t^2\partial_t+tu\partial_u,$\\
&  & & & $ \alpha t\partial_t + 2[-3\beta x^2+\gamma_1 x+\gamma_0]\partial_x+2\beta xu\partial_u$  \\
\thetbn b &  $-4$ & $u^{-4}$ & $f^4(x)|_{\mu=-4}$ &
$\partial_t,\,2t\partial_t+u\partial_u,\, t^2\partial_t+tu\partial_u,$\\
&  & & & $ \alpha t\partial_t -2[-3\beta e^{-x}+\gamma_1+\gamma_0 e^{x}]\partial_x+2\beta e^{-x}u\partial_u$  \\
\refstepcounter{tbn}\label{CaseHu-4K0F1} \thetbn a
& $-4$ & $0$ & $1$ &
$\partial_t,\, 2t\partial_t+u\partial_u,\,
t^2\partial_t+tu\partial_u,\, \partial_x,\, 2x\partial_x-u\partial_u$  \\
\thetbn b & $-4$ & $u^{-4}$ & $e^{-2x}$ &
$\partial_t,\, 2t\partial_t+u\partial_u,\, t^2\partial_t+tu\partial_u,\,
2\partial_x+u\partial_u,\, e^x\partial_x$  \\
\thetbn c &  $-4$ & $0$ & $x^{-\frac{8}{3}}$ &
$\partial_t,\, 2t\partial_t+u\partial_u,\, t^2\partial_t+tu\partial_u,\,
6x\p_x+u\p_u,\, 3x^2\partial_x-xu\partial_u$  \\
\thetbn d &  $-4$ & $u^{-4}$ &
$\frac{e^{-2x}}{(e^{-x}+\gamma)^{\frac{8}{3}}}$ &
$\partial_t,\, 2t\partial_t+u\partial_u,\, t^2\partial_t+tu\partial_u,\,
t\partial_t +3(e^{-x}+\gamma)e^{x}\partial_x,$\\
&  & & & $3(e^{-x}+\gamma)^2e^x\partial_x+(e^{-x}+\gamma)u\partial_u$   \\
\refstepcounter{tbn}\label{CaseHu-43K0F1} \thetbn a
& $-\frac{4}{3}$ & $0$ & $1$ &
$\partial_t,\, 2t\partial_t+3u\partial_u,\, \partial_x,\, 2x\partial_x-3u\partial_u, x^2\partial_x-3xu\partial_u$  \\
\thetbn b & $-\frac{4}{3}$ & $u^{-\frac{4}{3}}$ & $e^{-2x}$ &
$\partial_t,\,2t\partial_t+3u\partial_u,\,2\partial_x+3u\partial_u,\, e^{-x}(\partial_x+3u\partial_u),\,e^x\partial_x$\\
\hline
\end{tabular}
\end{center}
{\footnotesize Here $(\mu,\nu)\ne(0,0)$, $\lambda\neq 0$, $\epsilon\in \{0, 1\}\!\!\mod G^{\sim}$;
 $\mu \neq 0$ for cases~\ref{CaseHpowerK1Fpower}--\ref{CaseHpowerK0F1}d;
$\alpha, \zeta_0, \beta, \gamma_1, \gamma_0=$ const and
\[
f^3(x)=\exp\left\{\int \frac{-(3\mu+4)\beta
x-2\gamma_1+\alpha}{(\mu+1)\beta x^2+\gamma_1 x+\gamma_0}
dx\right\},\qquad
f^4(x)=\exp\left\{\int \frac{(\mu+2)\beta
e^{-x}-\alpha-2\gamma_0e^x}{(\mu+1)\beta
e^{-x}+\gamma_1+\gamma_0e^x} dx\right\}.
\]
Additional equivalence transformations:\\ \setcounter{casetran}{0}
\refstepcounter{casetran}\thecasetran.
 \ref{CaseHpowerK0Ff3}b $\to$ \ref{CaseHpowerK0Ff3}a, \ref{CaseHpowerK0F1}b$\to$\ref{CaseHpowerK0F1}a,
 \ref{CaseHu-4K0Ff3}b $\to$ \ref{CaseHu-4K0Ff3}a, \ref{CaseHu-4K0F1}b $\to$ \ref{CaseHu-4K0F1}a,
 \ref{CaseHu-43K0F1}b $\to$ \ref{CaseHu-43K0F1}a :\quad
    $\tilde t=t$, $\tilde x=e^{-x}$, $\tilde u=u;$\\
\refstepcounter{casetran}\thecasetran.
 \ref{CaseHpowerK0F1}c $\to$ \ref{CaseHpowerK0F1}a, \ref{CaseHu-4K0F1}c $\to$ \ref{CaseHu-4K0F1}a ($\mu=-4$):\quad
    $ \tilde t=t$, $\tilde x=-\frac{1}{x}$, $\tilde u=|x|^{-\frac{1}{1+\mu}}u.$\\
\refstepcounter{casetran}\thecasetran.
 \ref{CaseHpowerK0F1}d $\to$ \ref{CaseHpowerK0F1}a, \ref{CaseHu-4K0F1}d $\to$ \ref{CaseHu-4K0F1}a ($\mu=-4)$:\quad
   $ \tilde t=t$, $\tilde x=-\frac{1}{\gamma+e^{-x}}$, $\tilde u=|\gamma+e^{-x}|^{-\frac{1}{1+\mu}}u.$\\
\refstepcounter{casetran}\thecasetran. \ref{CaseHpowerK0F1}e $\to$
\ref{CaseHpowerK0F1}a ($\mu=-1$):\quad
   $ \tilde t=t$, $\tilde x=x$, $\tilde u=e^{\gamma x}u;$\\
\refstepcounter{casetran}\thecasetran. \ref{CaseHpowerK0F1}f $\to$
\ref{CaseHpowerK0F1}a ($\mu=-1$):\quad
   $ \tilde t=t$, $\tilde x=e^{-x}$, $\tilde u=e^{-\gamma
   e^{-x}}u.$\\
\refstepcounter{casetran}\thecasetran. \ref{CaseHpowerK0F1}g $\to$
\ref{CaseHpowerK0F1}a ($\mu=-2$):\quad $\tilde t=t,$ $\tilde x=e^x,$ $\tilde u=e^{-x}u.$
}

\begin{proof}
To prove the theorem we use the compatibility
method~\cite{Nikitin&Popovych2001,Popovych&Ivanova2004NVCDCEs}.
The basic idea of this method is based on the fact that the
substitution of the coefficients of any operator from \mbox{$A^{\max}\setminus A^{\ker}$}
into the classifying equations results in
nonidentity equations for arbitrary elements
(see~\cite{Nikitin&Popovych2001,Popovych&Ivanova2004NVCDCEs} for
more details and exhaustive examples of applications).
In our case the procedure of looking for the possible cases
mostly depends on equation~\eqref{sysDetEqTelEq_2}. For any
symmetry operator equation~\eqref{sysDetEqTelEq_2} gives
some equations on $H$ of the general form
$(au+b)H_u=cH,$
where $a$, $b$, $c$ are constant. For all operators from
$A^{\max}$ the number $k$ of such independent equations is not
greater than $2$; otherwise they form an incompatible system on
$H$. $k$ is an invariant value for the transformations from
$G^{\sim}$. Therefore, there exist three inequivalent cases for
the value of $k$:
i) $k = 0$: $H(u)$ is arbitrary; ii) $k = 1$: $H(u)=e^{u}$
or $H(u)=u^{\mu}$ $(\mu\neq 0)\!\! \mod G^{\sim}$,
and iii) $k = 2$: $H(u)=1\!\!\mod G^{\sim}$.
Let us consider in more detail case $H(u)=e^{u}$ (table~\ref{TableGrClasHexp}).
We attempted to present our calculations in reasonable
detail so that verification would be feasible.
For this case equations~\eqref{sysDetEqTelEq_2} and~\eqref{CoefsSymIntegr} imply $\eta_u=0$, i.e.
$\eta=\eta(t, x)$ and $\tau_{tt}=0$. Therefore, equations~\eqref{sysDetEqTelEq_2}--\eqref{sysDetEqTelEq_4}
can be written as
\begin{gather}
2(\xi_x-\tau_t)+\frac{f_x}{f}\xi=\eta,\label{sysDetEqHexp_1}\\
e^{u}\eta_{xx}+K\eta_x-f\eta_{tt}=0,\label{sysDetEqHexp_2}\\
(K-K_u)\eta-K\xi_x-2e^{u}\eta_x+e^{u}\xi_{xx}=0.\label{sysDetEqHexp_3}
\end{gather}
Equation~\eqref{sysDetEqHexp_3} looks like $K_u=\nu K+be^{u}$ with
respect to $K$, where $\nu, b=\const$. Therefore, $K$ must take
one of the following four values.

(i) $K=e^{\nu u}+h_1e^u$ mod $G^{\sim}$, where $\nu\bar\in\{0,
1\}$, $h_1=\const$. Substituting $K$ into
equations~\eqref{sysDetEqHexp_2} and~\eqref{sysDetEqHexp_3} yields
$\eta=\const$, $h_1=0$ and $\xi_x=(\mu-\nu)\eta$ which implies
there exist two cases for $f$, i.e. $f\neq\const$ or not by
further considering equation~\eqref{sysDetEqHexp_1}. Thus
we get cases~\ref{CaseHexpKexpnuuFxlambda} and~\ref{CaseHexpKexpnuuF1}.

(ii) $K=ue^{u}+h_1e^u\mod G^{\sim}$ and $h_1=\const$. It follows
from equations~\eqref{sysDetEqHexp_2} and~\eqref{sysDetEqHexp_3} that $\eta=0$ for any operator from
$A^{\max}$, which contradict with the assumption that $\eta\not\equiv 0$.

(iii) $K=e^{u}+h_0\!\!\mod G^{\sim}$, where $h_0=\const$.
Substituting $K$ into~\eqref{sysDetEqHexp_2} and~\eqref{sysDetEqHexp_3} yields
\begin{gather}
\eta_{xx}+\eta_x=0, \quad h_0\eta_x-f\eta_{tt}=0,\quad
\xi_{xx}-\xi_x-2\eta_x=0, \quad h_0(\eta-\xi_x)=0.\label{sysDetEqHexp_4}
\end{gather}
Solving the first and the third equation of system~\eqref{sysDetEqHexp_4} we obtain
$\eta=\zeta^1(t)e^{-x}+\zeta^0(t)$,
$\xi=\gamma_0e^x+\gamma_1-\zeta^1(t)e^{-x}$. Since $\xi_t=0$, we
have $\zeta^1_t=0$. Then it follows from the second and the fourth
equation of system~\eqref{sysDetEqHexp_4} and equation~\eqref{sysDetEqHexp_1} that $h_0=0$ and
\[
\tau=\frac{1}{2}(c_2-\alpha)t+c_1,\quad\quad
\xi=\gamma_0e^x+\gamma_1+\beta e^{-x},\quad\quad \eta=-\beta
e^{-x}-c_2,
\]
where $c_1$, $c_2$, $\alpha$, $\beta$, $\gamma_0$, $\gamma_1=\const$.
Hence, equation~\eqref{sysDetEqHexp_1} implies that the function
$f$ must satisfy $l$ $(l = 0, 1, 2)$ equations of the form
\[
\frac{f_x}{f}=\frac{\beta e^{-x}-\alpha-2\gamma_0e^x}{\beta e^{-x}+\gamma_1+\gamma_0e^x}
\]
with non-proportional sets of constant parameters $(\alpha, \beta,
\gamma_0, \gamma_1)$. The values $l = 0$ and $l = 1$ correspond to
cases \ref{CaseHexpK0expuForAllF} $(\epsilon = 1)$ and
\ref{CaseHexpK0Ff1}b. An additional extension of $A^{\max}$ exists
for $l = 2$ in comparison with $l = 1$ iff $f$ is a
solution of the equation
\[
\frac{f_x}{f}=\frac{\lambda_2 e^{-x}}{\lambda_1 e^{-x}+\lambda_0}-2,
\]
where either $\lambda_2=0$ or $\lambda_2=3\lambda_1\neq 0$.
Integrating the latter equation gives cases \ref{CaseHexpK0F1}b
and \ref{CaseHexpK0F1}d.

(iv) $K=h_0=\const\!\! \mod G^{\sim}$. In an analogous way to that in
the previous case, we obtain $\eta=\zeta^1(t)x+\zeta^0(t),
\xi=\gamma_1x+\gamma_0+\zeta^1(t)x^2$, what is more, $\xi$, $\eta$
satisfy
\[
h_0\eta_x-f\eta_{tt}=0, \qquad h_0(\eta-\xi_x)=0.
\]
Investigating the compatibility of the latter system and equation~\eqref{sysDetEqHexp_1}
with $\xi_t=0$ leads to
\[
\tau=\frac{1}{2}(c_2+\alpha)t+c_1,\quad
\xi=\beta x^2+\gamma_1x+\gamma_0,\quad \eta=\beta x-c_2,
\]
where $c_1$, $c_2$, $\alpha$, $\beta$, $\gamma_0$,
$\gamma_1=\const$, and $\beta$, $c_2$, $\gamma_1$ satisfy
$h_0\beta=0$, $h_0(c_2+\gamma_1)=0$. Hence, there exist two cases
for $h_0$, i.e. $h_0=0$ or not. The value $h_0=1\mod G^{\sim}$
results in cases \ref{CaseHexpK1Fxlambda}
 and \ref{CaseHexpK1F1}. Below, $h_0=0$.
Equation~\eqref{sysDetEqHexp_1} holds when the function $f$ is a
solution of a system of $l (l = 0, 1, 2)$ equations of the form
\[
\frac{f_x}{f}=\frac{-3\beta x+\alpha-2\gamma_1}{\beta x^2+\gamma_1x+\gamma_0}
\]
with non-proportional sets of constant parameters $(\alpha, \beta,
\gamma_0, \gamma_1)$. The values $l = 0$ and $l = 1$ correspond to
cases \ref{CaseHexpK0expuForAllF} $(\epsilon = 0)$ and
\ref{CaseHexpK0Ff1}a. Additional extension of $A^{\max}$ exists
for $l = 2$ in comparison with $l = 1$ if and only if $f$ is a
solution of the equation
\[
\frac{f_x}{f}=\frac{\lambda_2 }{\lambda_1x+\lambda_0}
\]
where either $\lambda_2=0$ or
$\lambda_2=-3\lambda_1\neq 0$. These possibilities result in cases
\ref{CaseHexpK0F1}a and \ref{CaseHexpK0F1}c.

The rest of the cases of values $H$ can be studied in an analogous way.
\end{proof}

In what follows, for convenience we use double numeration $T.N$ of
classification cases and local equivalence transformations, where
$T$ denotes the number of the table and $N$ the number of the case
(or transformation) in table $T$. The notation `equation $T.N$' is
used for the equation of the form~\eqref{eqVarCoefTelegraphEq} where the parameter
functions take the values from the corresponding case.

The operators from
tables~\ref{TableGrClasForAllH}--\ref{TableGrClasHpower} form
bases of the maximal invariance algebras if the corresponding sets
of the functions $f$, $H$, $K$ are $G^{\sim}$-inequivalent to ones
with most extensive invariance algebras. For example, in case
$\ref{TableGrClasHpower}.\ref{CaseHpowerKpowerFpower}~ (\mu,
\nu)\neq(0, 0)$ and $\lambda \neq -6$ if $\nu =1$. Similarly, in
case \ref{TableGrClasHexp}.\ref{CaseHexpKexpnuuFxlambda} the
constraint set on the parameters $\mu, \nu$ and $\lambda$
coincides with the one for case
\ref{TableGrClasHpower}.\ref{CaseHpowerKpowerFpower}, and $\mu =
1$ if $\nu = 0$.

\section{Additional equivalence transformations and
classification \\with respect to the set of point transformations}\label{SectionOnAdditEquivTr}

In tables~\ref{TableGrClasForAllH}--\ref{TableGrClasHpower} we
list all possible $G^{\sim}$-inequivalent sets of functions
$f(x)$, $H(u)$, $K(u)$ and corresponding invariance algebras.
However, these tables contain some cases being equivalent with respect to
point transformations that do not belong to~$G^{\sim}$.
The simplest way to find such additional equivalences between previously
classified equations is based on the fact that equivalent
equations have equivalent maximal Lie invariance algebras.

Explicit formulas for additional transformations that do not change the value of $H(u)$ are
adduced after the tables.
Besides these transformations there exist additional point transformations
changing~$H(u)$. Thus, e.g.,
\[
\tilde t=x,\quad \tilde x=t,\quad \tilde u=\ln u
\]
maps case~\ref{TableGrClasHpower}.\ref{CaseHpowerK0F1}a to \ref{TableGrClasHexp}.\ref{CaseHexpK0F1}a.
One more example of similar transformations is
\[
\tilde t=x,\quad \tilde x=t,\quad \tilde u=u^{\mu+1},\quad \tilde\mu=-\mu/(\mu+1)
\]
between equations of form $u_{tt}=(u^\mu u_x)_x$, $\mu\ne-1$.  In particular,
it connects cases~\ref{TableGrClasHpower}.\ref{CaseHu-4K0F1}a and~\ref{TableGrClasHpower}.\ref{CaseHu-43K0F1}a.
The same transformation applied is a discrete symmetry for equation with $\mu=-2$.
The latter two transformations are, indeed, partial cases of more general transformation
\begin{equation}\label{TrExtEquivTransf}
\tilde t=x,\quad \tilde x=t,\quad \tilde u=\int H(u)du
\end{equation}
between equations from class
\begin{equation}\label{eqWaveEq}
u_{tt}=(H(u) u_x)_x,
\end{equation}
where the new transformed value of arbitrary element $\tilde H$
is the derivative to the inverse function $u=\hat H(\tilde u)$ for $\tilde u=\int H(u)du$~\cite{Ibragimov1994V1}.
Transformation~\eqref{TrExtEquivTransf} is nonlocal with respect to the arbitrary element~$H(u)$
and therefore can be considered as
{\em generalized extended equivalence transformation}~\cite{Meleshko1994,Ivanova&Popovych&Sophocleous2004}
in class of nonlinear wave equations~\eqref{eqWaveEq}.

One can check that there exist no other point transformations between the equations from
tables~\ref{TableGrClasForAllH}--\ref{TableGrClasHpower}.
Using this we can formulate the following theorem.

\begin{theorem}\label{TheoremOnClassificationOfDCEfghWRTPointTrans}
Up to point transformations, a complete list of extensions of the maximal
Lie invariance algebra of equations from class~\eqref{eqVarCoefTelegraphEq}
is exhausted by the cases given in table~\ref{TableGrClasForAllH},
cases \ref{TableGrClasHexp}.1--\ref{TableGrClasHexp}.\ref{CaseHexpK0Ff1}a
and \ref{TableGrClasHpower}.1--\ref{TableGrClasHpower}.\ref{CaseHu-4K0F1}a
numbered with Arabic numbers without Roman letters and
subcases ``a'' of each multi-case.
(Two equations from case~\ref{TableGrClasHpower}.\ref{CaseHpowerK0F1}
with parameter values $\mu$ and $\tilde\mu$ are assumed to be equivalent iff
$\tilde\mu=-\mu/(\mu+1)$).
\end{theorem}

As one can see, the above additional equivalence transformations have multifarious structure.
This displays a complexity of a structure of the set of admissible transformations.
Usually the problems of finding of all possible admissible transformations are very difficult to solve,
see, e.g.,~\cite{Kingston&Sophocleous1998,Kingston&Sophocleous2001,
Popovych&Ivanova&Eshraghi2004Gamma,Popovych2006}.
We will try to discuss the structure of the set of admissible
transformations of class~\eqref{eqVarCoefTelegraphEq} in a sequel paper.

A more systematic way to proceed with equivalence transformations is to classify them
using the infinitesimal method or the direct method. Examples of
conditional equivalence algebras calculated by the infinitesimal
method are listed in table~\ref{TableCondEquivAlg}.

\setcounter{tbn}{0}

\begin{center}\footnotesize\renewcommand{\arraystretch}{1.15}
Table~\refstepcounter{table}\label{TableCondEquivAlg}\thetable. Conditional equivalence algebras \\[1ex]
\begin{tabular}{|l|l|}
\hline
Conditions   & \hfill  Basis of A$^{\max}\hfill $ \\
\hline \refstepcounter{tbn}\label{CondiEquivalge5} $K=H$ &
$\partial_t,\, \partial_x,\, t\partial_t+2f \partial_f,\,
\partial_u,\, u\partial_u,\,
e^{x}( \partial_x-2f \partial_f),\, f \partial_f+H \partial_H$
\\
$K=H=e^u$  & $\partial_t,\, \partial_x,\, t\partial_t+2f\partial_f,\,
\partial_u+f\partial_f,\,
e^{x}( \partial_x-2f \partial_f),\, e^{-x}(-\partial_x+\partial_u-f\partial_f)$
\\
$H=e^u$, $K=0$  & $\partial_t,\, \partial_x,\, t\partial_t+2f\partial_f,\,
\partial_u+f\partial_f,\,
x\partial_x-2f \partial_f,\, x^2\partial_x+x\partial_u-3xf\partial_f$
\\
$K=H=u^{\mu}$  & $\partial_t,\, \partial_x,\, t\partial_t+2f\partial_f,\,
u\partial_u+\mu f\partial_f,\,
 e^{x}( \partial_x-2f \partial_f),\, e^{-x}[(1+\mu)\partial_x-u\partial_u+(2+\mu)f\partial_f]$
 \\
$H=u^{\mu}$, $K=0$  & $\partial_t,\, \partial_x,\, t\partial_t+2f\partial_f,\,
u\partial_u+\mu f\partial_f,\,
 x\partial_x-2f \partial_f,\, (1+\mu)x^2\partial_x+xu\partial_u-(4+3\mu)xf\partial_f$
 \\
\hline
\end{tabular}
\end{center}

\bigskip

To find the complete collection of additional local equivalence
transformations including both continuous and discrete ones, we
should use the direct method. Moreover, application of this method
allows us to describe all the local transformations that are
possible for pairs of equations from the class under
consideration. A problem of this sort was first investigated for
wave equations by Kingston and Sophocleous~\cite{Kingston&Sophocleous1998,Kingston&Sophocleous2001,
Sophocleous&Kingston1999}.

\begin{remark}
It is a well-known that class~\eqref{eqWaveEq} is linearizable~\cite{Bluman&Kumei1987}
with respect to potential hodograph transformation
(e.g., interchange of independent and dependent variables)  applied to the potential system
$v_x=u_t$, $v_t=H(u)u_x$ corresponding to the simplest local conservation law of~\eqref{eqWaveEq}.
This transformation can be considered as potential equivalence transformation between the classes of wave equations
$u_{tt}=F(x)u_{xx}$ and~\eqref{eqWaveEq}.
\end{remark}


\section{Exact solutions}\label{SectionOnExactSol}

In this section, we turn to the presentation of some exact
solutions for~\eqref{eqVarCoefTelegraphEq} by means of the
classical Lie--Ovsiannikov algorithm and generalized conditional
symmetry methods. We first present the solutions of some special
forms of the nonlinear wave equation~\eqref{eqWaveEq}.
Then using our classification with respect to all the possible local
transformations, we transform them to solutions of more complicated
telegraph equations (such as \ref{TableGrClasHexp}.\ref {CaseHexpK0F1}b,
\ref{TableGrClasHpower}.\ref{CaseHpowerK0F1}b,
\ref{TableGrClasHpower}.\ref{CaseHpowerK0F1}e).

\subsection{Exact solutions obtained via classical Lie--Ovsiannikov
algorithm}

Let us note that the equations with $f=1$ are well investigated
and that most of the exact solutions given below have been
constructed before (see citations
in~\cite{Bluman&Temuerchaolu&Sahadevan2005,Ibragimov1994V1,Kingston&Sophocleous2001,Gandarias&Torrisi&Valenti2004}).
However, to the best of our knowledge, there exist no works
containing a systematic study of all the possible Lie reductions
in this class, as well as exhaustive consideration of the
integrability and exact solutions of the corresponding reduced
equations. That is why we have decided to implement the relevant
Lie reduction algorithm independently, especially since it is not
a difficult problem.

So, let us consider equation
\ref{TableGrClasHexp}.\ref{CaseHexpK0F1}a:
\begin{equation}\label{eqWaveEqexp}
u_{tt}=(e^uu_x)_x.
\end{equation}
Let us recall that for~\eqref{eqWaveEqexp} the basis of $A^{\max}$
is formed by the operators
\[
Q_1=\partial_t,\quad Q_2=t\partial_t-\partial_u,\quad
Q_3=\partial_x,\quad Q_4=x\partial_x +2\partial_u.
\]
The only non-zero commutators of these operators are
$[Q_1,Q_2]=Q_1$ and $[Q_3, Q_4]=Q_3$. Therefore $A^{\max}$ is a
realization of the algebra $2A_{2.1}$~\cite{Mubarakzyanov1963}.
All the possible inequivalent (with respect to inner
automorphisms) one-dimensional subalgebras of
$2A_{2.1}$~\cite{Patera&Winternitz1977} are exhausted by the ones
listed in table~\ref{TableRedODEsWaveEqexp} along with the
corresponding ans\"{a}tze and the reduced ODEs.

\setcounter{tbn}{0}

\begin{center}\footnotesize\renewcommand{\arraystretch}{1.1}
Table~\refstepcounter{table}\label{TableRedODEsWaveEqexp}\thetable.
Reduced ODEs for equation~\eqref{eqWaveEqexp}.
$\alpha\neq 0$, $\epsilon=\pm 1$. \\[1ex]
\begin{tabular}{|l|c|c|c|l|}
\hline
N & Subalgebra & Ansatz $u=$ & $\omega$ & \hfil Reduced ODE \\
\hline \refstepcounter{tbn}\label{ReduODEForCaseHexpK0F11}\thetbn
& $\langle Q_1\rangle $ & $\varphi(\omega)$ & $x$
& $(e^{\varphi})''=0$  \\
\refstepcounter{tbn}\label{ReduODEForCaseHexpK0F12}\thetbn &
$\langle Q_2\rangle $ & $\varphi(\omega)-2\ln|t|$ & $x$
& $(e^{\varphi})''=2$  \\
\refstepcounter{tbn}\label{ReduODEForCaseHexpK0F13}\thetbn &
$\langle Q_3\rangle $ & $\varphi(\omega)$ & $t$
& $\varphi''=0$  \\
\refstepcounter{tbn}\label{ReduODEForCaseHexpK0F14}\thetbn &
$\langle Q_4\rangle $ & $\varphi(\omega)+2\ln|x|$ & $t$
& $(e^{\varphi})''=2e^{\varphi}$  \\
\refstepcounter{tbn}\label{ReduODEForCaseHexpK0F15}\thetbn &
$\langle Q_1+\epsilon Q_3\rangle $ & $\varphi(\omega)$ &
$x-\epsilon t$
& $(e^{\varphi})''=\epsilon^2\varphi''$   \\
\refstepcounter{tbn}\label{ReduODEForCaseHexpK0F16}\thetbn &
$\langle Q_2+\epsilon Q_3\rangle $ & $\varphi(\omega)-2\ln|t|$ &
$x-\epsilon \ln|t|$
& $(e^{\varphi})''=\epsilon^2\varphi''+\epsilon\varphi'+2$  \\
\refstepcounter{tbn}\label{ReduODEForCaseHexpK0F17}\thetbn &
$\langle Q_1+\epsilon Q_4\rangle $ & $\varphi(\omega)+2\epsilon t$
& $xe^{-\epsilon t}$
& $(e^{\varphi})''=\epsilon^2\omega(\omega\varphi''+\varphi')$  \\
\refstepcounter{tbn}\label{ReduODEForCaseHexpK0F18}\thetbn &
$\langle Q_2+\alpha Q_4\rangle $ &
$\varphi(\omega)+2(\alpha-1)\ln|t|$ & $x|t|^{-\alpha}$
& $(e^{\varphi})''=\alpha^2\omega^2\varphi''+\alpha(\alpha+1)\omega\varphi'-2(\alpha-1)$  \\
\hline
\end{tabular}
\end{center}

We succeeded in solving the equations
\ref{TableRedODEsWaveEqexp}.\ref{ReduODEForCaseHexpK0F11}--\ref{TableRedODEsWaveEqexp}.\ref{ReduODEForCaseHexpK0F15}.
Thus we have the following solutions of~\eqref{eqWaveEqexp}:
\begin{gather*}
u=\ln|c_1x+c_0|,\quad
u=\ln\left|\frac{x^2+c_1x+c_0}{t^2}\right|,\quad
u=c_1t+c_0,\\
 u=\ln\Big(\frac{1}{4c_0^2\cosh^2(\frac{t+c_1}{2c_0})}+x^2\Big),\quad
 u=\varphi(x-\epsilon t),
\end{gather*}
where $\varphi$ satisfies
$e^{\varphi}=\epsilon^2\varphi+c_1\varphi+c_0.$ Using these we can
construct solutions for cases \ref{TableGrClasHexp}.\ref{CaseHexpK0F1}b--\ref{TableGrClasHexp}.\ref{CaseHexpK0F1}d
easily. For example, the transformation 3.4 yields the
corresponding solutions for the more complicated and interesting
equation (case \ref{TableGrClasHexp} .\ref{CaseHexpK0F1}d)
\begin{equation}\label{EqVarCoeExpWave}
e^{-2x}(e^{-x}+\gamma)^{-3}u_{tt}=(e^uu_x)_x+e^uu_x
\end{equation}
in the following form
\begin{gather*}
u=\ln\mid c_1x+c_0(e^{-x}+\gamma)|,\quad
u=\ln\Big|\frac{1}{t^2(e^{-x}+\gamma)}+\frac{c_1}{t^2}+\frac{c_0}{t^2}(e^{-x}+\gamma)\Big|,\quad
u=c_1t+c_0.
\end{gather*}

The power $\mu=-1$ is a singular value of the parameter $\mu$ for case
\ref{TableGrClasHpower}.\ref{CaseHpowerK0F1}a. So, the corresponding equation
\begin{equation}\label{eqWaveEqpower-1}
u_{tt}=(u^{-1}u_x)_x
\end{equation}
is distinguished by the reduction procedure. It is remarkable that
cases \ref{TableGrClasHpower}.\ref{CaseHpowerK0F1}e and
\ref{TableGrClasHpower}.\ref{CaseHpowerK0F1}f are reduced exactly
to equation~\eqref{eqWaveEqpower-1}.
Exact solutions of equation~\eqref{eqWaveEqpower-1} can be easily obtained
by direct application of the classical Lie reduction method
or using transformation~\eqref{TrExtEquivTransf} applied to solutions of equation~\eqref{eqWaveEqexp}.
Both these approaches lead us to the following solutions of~\eqref{eqWaveEqpower-1}:
\begin{gather*}
u=c_2e^{c_1x},\quad
u=-\frac{t^2}{4c_1^2\cosh^2(\frac{x+c_2}{2c_1})},\quad u=c_1t+c_0,\quad
u=\frac{t^2}{4c_1^2\cos^2(\frac{x+c_2}{2c_1})},\\
u=x^{-2}(t^2+c_2t+c_1),\quad u=\varphi(x-\epsilon t),
\end{gather*}
where $\varphi$ satisfies$\int
\frac{1-\varphi}{c_2\varphi}d\varphi=\omega+c_1$. Analogously to
the previous case, we obtain by means of transformations 4.5 exact
solutions of equation
\ref{TableGrClasHpower}.\ref{CaseHpowerK0F1}f, i.e.
\begin{equation}\label{eqVarCoefWaveEqpower-1}
e^{-2x+\gamma e^{-x}}u_{tt}=(u^{-1}u_x)_x+u^{-1}u_x
\end{equation}
in the following forms:
\begin{gather*}
u=c_2e^{(c_1+\gamma)e^{-x}},\quad u=-\frac{t^2e^{\gamma
e^{-x}}}{4c_1^2\cosh^2(\frac{e^{-x}+c_2}{2c_1})},\quad u=c_1t+c_0,\quad
u=\frac{t^2e^{\gamma e^{-x}}}{4c_1^2\cos^2(\frac{e^{-x}+c_2}{2c_1})},\\
u=e^{-2x+\gamma e^{-x}}(t^2+c_2t+c_1),\quad u=e^{\gamma
e^{-x}}\varphi(e^{-x}-\epsilon t),
\end{gather*}
where $\varphi$ is as above.

Another example of a variable coefficient equation 
is given by case \ref{TableGrClasHpower}.\ref{CaseHpowerK0F1}d. To
look for exact solutions of it, first we reduce it to equation
\begin{equation}\label{eqWaveEqpower}
u_{tt}=(u^{\mu}u_x)_x
\end{equation}
(case \ref{TableGrClasHpower}.\ref{CaseHpowerK0F1}a).
As in the previous cases, the invariance algebra
of~\eqref{eqWaveEqpower} is of the form
\[
Q_1=\partial_t,\quad Q_2=t\partial_t-2\mu^{-1}u\partial_u,\quad
Q_3=\partial_x,\quad Q_4=x\partial_x +2\mu^{-1}u\partial_u.
\]
It is also a realization of the algebra $2A_{2.1}$. The reduced
ODEs for equation~\eqref{eqWaveEqpower} are listed in
table~\ref{TableRedODEsWaveEqpower}.

\setcounter{tbn}{0}

\begin{center}\footnotesize\renewcommand{\arraystretch}{1.15}
Table~\refstepcounter{table}\label{TableRedODEsWaveEqpower}\thetable.
 Reduced ODEs for equation~\eqref{eqWaveEqpower}. $\alpha\neq 0, \epsilon=\pm 1$. \\[1ex]
\begin{tabular}{|l|c|c|c|l|}
\hline
N & Subalgebra & Ansatz $u=$ & $\omega$ & \hfil Reduced ODE \\
\hline \refstepcounter{tbn}\label{ReduODEFor81}\thetbn & $\langle
Q_1\rangle $ & $\varphi(\omega)$ & $x$
& $(\varphi^{\mu}\varphi')'=0$  \\
\refstepcounter{tbn}\label{ReduODEFor82}\thetbn & $\langle
Q_2\rangle $ & $\varphi(\omega)|t|^{-\frac{2}{\mu}}$ & $x$
& $(\varphi^{\mu}\varphi')'=\frac{2}{\mu}(\frac{2}{\mu}+1)\varphi$  \\
\refstepcounter{tbn}\label{ReduODEFor83}\thetbn & $\langle
Q_3\rangle $ & $\varphi(\omega)$ & $t$
& $\varphi''=0$  \\
\refstepcounter{tbn}\label{ReduODEFor84}\thetbn & $\langle
Q_4\rangle $ & $\varphi(\omega)|x|^{\frac{2}{\mu}}$ & $t$
& $\varphi''=2\mu^{-2}(\mu+2)\varphi^{\mu+1}$  \\
\refstepcounter{tbn}\label{ReduODEFor85}\thetbn & $\langle
Q_1+\epsilon Q_3\rangle $ & $\varphi(\omega)$ & $x-\epsilon t$
& $(\varphi^{\mu}\varphi')'=\epsilon^2\varphi''$   \\
\refstepcounter{tbn}\label{ReduODEFor86}\thetbn & $\langle
Q_2+\epsilon Q_3\rangle $ & $\varphi(\omega)|t|^{-\frac{2}{\mu}}$
& $x-\epsilon \ln|t|$ &
$(\varphi^{\mu}\varphi')'=\epsilon^2\varphi''+\epsilon(\frac{4}{\mu}+1)\varphi'
+\frac{2}{\mu}(\frac{2}{\mu}+1)\varphi$  \\
\refstepcounter{tbn}\label{ReduODEFor87}\thetbn & $\langle
Q_1+\epsilon Q_4\rangle $ & $\varphi(\omega)e^{2\epsilon\mu^{-1}
t}$ & $xe^{-\epsilon t}$ &
$(\varphi^{\mu}\varphi')'=\epsilon^2\omega^2\varphi''+4\epsilon^2\mu^{-2}\varphi
-(4\mu^{-1}-1)\epsilon^2\omega\varphi'$  \\
\refstepcounter{tbn}\label{ReduODEFor88}\thetbn & $\langle
Q_2+\alpha Q_4\rangle $ & $\varphi(\omega)|
t|^{\frac{2(\alpha-1)}{\mu}}$ & $x|t|^{-\alpha}$ &
$(\varphi^{\mu}\varphi')'=\alpha^2\omega^2\varphi''-\alpha[(4\alpha-4)\mu^{-1}-\alpha-1]\omega\varphi'$
\\ & ~ & ~ & ~ & $+2(\alpha-1)\mu^{-1}[(2\alpha-2)\mu^{-1}-1]\varphi$  \\
\hline
\end{tabular}
\end{center}

For some of the reduced equations we can construct the general
solutions. For others we succeeded in finding only particular
solutions. These solutions are the following:
\[
u=|c_2-c_1x-c_1 \mu x|^{\frac{1}{1+\mu}},\quad u=c_1t+c_0,\quad
u=\varphi(x-\epsilon t),
\]
where $\varphi$ satisfies
$\frac{1}{1+\mu}\varphi^{1+\mu}-\varphi-c_1\omega-c_2=0.$ All the
results of table~\ref{TableRedODEsWaveEqpower}
as well as the solutions constructed can be extended to equations
\ref{TableGrClasHpower}.\ref{CaseHpowerK0F1}b--\ref{TableGrClasHpower}.\ref{CaseHpowerK0F1}f
using the local equivalence transformations. So for the equation
(case \ref{TableGrClasHpower}.\ref{CaseHpowerK0F1}d)
\[
e^{-2x}(e^{-x}+\gamma)^{-\frac{3\mu+4}{\mu+1}}u_{tt}=(u^{\mu}u_x)_x+u^{\mu}u_x
\]
the transformations 4.3 yield exact solutions in the form
\[
u=|c_2(e^{-x}+\gamma)+c_1+c_1 \mu |^{\frac{1}{1+\mu}},\quad u=c_1t+c_0,\quad
u=|
(e^{-x}+\gamma)|^{\frac{1}{1+\mu}}\varphi(-(e^{-x}+\gamma)^{-1}-\epsilon t),
\]
with the same value of $\varphi$.

\subsection{Functionally separation solutions obtained via generalized conditional
symmetry methods}

We now turn to the functionally separation solutions of the
nonlinear telegraph equations~\eqref{eqConstCoefTelegraphEq} by
using the generalized conditional symmetry
methods~\cite{Fokas&Liu1994,Zhdanov1995,Qu1997}. This method was
developed by Fokas, Zhdanov and Qu {\it et al}, and has been applied
to study the functional separation of variables for various
nonlinear equation \cite{Qu&He&Dou2001,Qu&Zhang&Liu2000}. In order
to implement the method effectively, let us review some basic
notations of the functionally separation solutions and generalized
conditional symmetry~\cite{Qu1997,Qu&He&Dou2001,Qu&Zhang&Liu2000}.

\begin{definition}\label{DefFunctSepaSolu}
A solution $u(t,x)$ of equation~\eqref{eqConstCoefTelegraphEq} is
said to be functionally separable if there exist functions $q(u)$,
$\varphi(t)$, and $\psi(x)$ such that
\begin{equation}\label{ExpreFunctSepaSolu}
q(u) = \varphi(t) + \psi(x),
\end{equation}
where $q(u)$ is some smooth function of $u$, $\varphi(t)$ and
$\psi(x)$ are some undetermined functions of $t$ and $x$
respectively.
\end{definition}
The classical additively separable solution and product separable
solution are particular cases of the above functional separable
solution.

\begin{definition}\label{DefGeneCondSymm}
An evolutionary vector field
\[
V = \eta(t,x,u,\ldots)\partial_u
\]
is said to be a generalized conditional symmetry of~\eqref{eqConstCoefTelegraphEq} if
\begin{equation}\label{ProloDeterequation}
\pr V^{(2)}(u_{tt}-(H(u)u_x)_x-K(u)u_x)|_{E\cap W} = 0,
\end{equation}
where $E$ is the solution manifold of~\eqref{eqConstCoefTelegraphEq} and $W$ is a second-order system
of~\eqref{eqConstCoefTelegraphEq} obtained by appending the
condition $\eta= 0$ and its differential consequences; $\pr V^{(2)}$ is the second
prolongation of the infinitesimal operator $V$.
\end{definition}

According to reference~\cite{Qu&He&Dou2001}, there exist the
following theorem.

\begin{theorem}\label{TheoremOnExpresGeneCondSymm}
Equation~\eqref{eqConstCoefTelegraphEq} possesses the
functional separable solution~\eqref{ExpreFunctSepaSolu} iff
it admits the generalized conditional symmetry
\begin{equation}\label{ExpresEvoluVectorField}
V = (u_{xt}+g(u)u_xu_t)\partial_u,\quad g=\frac{q''(u)}{q'(u)}.
\end{equation}
\end{theorem}

Substituting~\eqref{ExpresEvoluVectorField} into~\eqref{ProloDeterequation}
and using~\eqref{eqConstCoefTelegraphEq}, a straightforward calculation
shows that the functions $H$, $K$ and $g$ satisfy the following
system
\begin{equation}\label{GreConSymmsysDetEqTelEqComp}
\begin{array}{ll}
H''-gH'-\frac{H'^2}{H}=0,\quad
K'-K\frac{H'}{H}=0,\quad
g''-2gg'+(g^2-g')\frac{H'}{H}=0,\\[0.5ex]
(2gg'-g'')H+(g^2-g')H'-2gH''-2g'H'+H'''+gH''\\[0.5ex]
-(gH'+H'')\frac{H'}{H}+H'(3g'-2g^2)=0,\\[0.5ex]
K''-g'K-(gK+K')\frac{H'}{H}+K(3g'-2g^2)=0.
\end{array}
\end{equation}

To obtain solutions of this system, we consider three cases for
the third equation.

{\bf (A) } $g'-g^2=0$. In this case, $g$ is given by $g=0$ and
$g=-1/u$ after by translating $u$. Substituting $g=0$ into the
remain equations of system~\eqref{GreConSymmsysDetEqTelEqComp}
and scaling $u$, we find that
\[
H=e^{\alpha u},\quad K=ce^{\alpha u},
\]
where $\alpha$, $c$ is arbitrary constant. We always assume
$\alpha\neq 0$ to exclude the cases
equation~\eqref{eqConstCoefTelegraphEq} is linear. Furthermore,
from~\eqref{ExpresEvoluVectorField}, we have $q=u$.

For $g=-1/u$, we can obtain
\[
H=u^{\alpha},\quad K=cu^{\alpha},\quad q=\ln u.
\]

{\bf (B) }  $g'-g^2\neq0$, $g''-2gg'=0$.
In this case, it is easy to see that $H$, $K=\const$.
Equation~\eqref{eqConstCoefTelegraphEq} is linear.
We thus do not consider it.

{\bf (C) } $g'-g^2\neq0$, $g''-2gg'\neq0$.
In order to solve system~\eqref{GreConSymmsysDetEqTelEqComp},
we define $h(u)$ such that $g=-{h'}/{h}$.
From the first and the third equation of system~\eqref{GreConSymmsysDetEqTelEqComp},
we obtain
\begin{equation}\label{GreConSymmExpreForH}
H=\frac{H_0h''}{h},\quad \frac{H'}{H}=\frac{a}{h},
\end{equation}
where $a$ is nonzero constant
(if $a=0$ equation~\eqref{eqConstCoefTelegraphEq} becomes linear),
$H_0$ is a constant and can be chosen as $\pm 1$ by scaling $t$;
$h$ satisfies
$
hh'''-h'h''=ah'',
$
which can be integrated as
\begin{equation}\label{GreConSymmEquaForh}
hh''-h'^2=ah'+b.
\end{equation}
Since $a\neq 0$, there exist two cases for solving equation~\eqref{GreConSymmEquaForh},
i.e., $b=0$ or not. If $b=0$, we can find from~\eqref{GreConSymmExpreForH},~\eqref{GreConSymmEquaForh}
and the rest equations of system~\eqref{GreConSymmsysDetEqTelEqComp} that
\[
h=\frac{a}{d}+ce^{du},\quad H=-\frac{H_0d^2e^{du}}{a/(dc)+e^{du}},\quad K=0,
\]
where $d$ is arbitrary constant. By scaling and translating $u$,
we can set $d=\pm 1$ and $a/c=\pm 1$. Hence, four possibilities
are distinguished:
\begin{gather*}
h=c(e^{-u}-1),\quad H=(e^u-1)^{-1},\quad K=0,\quad q=\ln(e^u-1),\\
\quad \mbox{when}\quad H_0=-d=1,
~a/c=1;\\
h=c(e^{u}+1),\quad H=e^u(e^u+1)^{-1},\quad K=0,\quad q=u-\ln(e^u+1),\\\quad \mbox{when}\quad H_0=-d=-1,
~a/c=1;\\
h=c(e^{u}-1),\quad H=e^u(e^u-1)^{-1},\quad K=0,\quad q=-u-\ln(e^u-1),\\\quad \mbox{when}\quad H_0=-d=-1,
~a/c=-1;\\
h=c(e^{-u}+1),\quad H=(e^u+1)^{-1},\quad K=0,\quad q=\ln(e^u+1),\\
\quad \mbox{when}\quad H_0=-d=-1, ~a/c=-1.
\end{gather*}

For $b\neq 0$ we introduce $F(h)=h_u$ in~\eqref{GreConSymmEquaForh}, $F$ satisfies
an ordinary differential equation $hFF_h=F^2+aF+b,$
which is integrated to
\begin{equation}\label{GreConSymmEquaWithF}
\int^F\frac{zdz}{z^2+az+b}=\ln\Big(\frac{h}{h_0}\Big),\quad  h_0=\mbox{const}.
\end{equation}
Set $\Delta=a^2-4b$, we must consider three cases separately for
the above integral.

For $\Delta=0$, we have
\[
\Big(h'+\frac{a}{2}\Big)e^{a/(2h'+a)}=\frac{h}{h_0},
\]
which can be rewritten as $F=h'=p_1(h)$.

For $\Delta>0$, we have
\[
\Big(h'+\frac{a+\sqrt{\Delta}}{2}\Big)^{\sqrt{\Delta}+a}
\Big(h'+\frac{a-\sqrt{\Delta}}{2}\Big)^{\sqrt{\Delta}-a}=\Big(\frac{h}{h_0}\Big)^{2\sqrt{\Delta}}.
\]
We rewrite it implicity as $h'=p_2(h)$.

For $\Delta<0$, we have
\[
\Big[1+\Big(\frac{2h'+a}{\sqrt{-\Delta}}\Big)^2\Big]\exp\Big[-2\frac{a}{\sqrt{-\Delta}}
\arctan\Big(\frac{2h'+a}{\sqrt{-\Delta}}\Big)\Big] =\Big(\frac{h}{h_0}\Big)^2.
\]
We rewrite it implicity as $h'=p_3(h)$.

In these cases, $h$, $H$ can be determined implicity by
\[
\int^h\frac{dz}{p_i(z)}=u,\quad H(u)=-\frac{p_i^2+ap_i+b}{h^2},\quad i=1,
2, 3,
\]
and $K$ satisfies the second and the fifth equations of
system~\eqref{GreConSymmsysDetEqTelEqComp}. The corresponding
equation has a separable solution of the form
\[
\int^h\frac{dz}{p_i(z)}=\varphi(t)+\psi(x).
\]

Summing up the above analysis, we have the following results:

\begin{theorem}\label{TheoremOnExpresOfEquWithFSV}
Equation~\eqref{eqConstCoefTelegraphEq} possesses the
functional separable solution~\eqref{ExpreFunctSepaSolu} if and
only if it is $G^{\sim}$ equivalent to one of the following
equations:
\begin{gather*}
{\rm(a)}\quad u_{tt}=(e^{u}u_x)_x+\epsilon e^{u}u_x,\quad u = \varphi(t) + \psi(x),\\
{\rm(b)}\quad u_{tt}=(u^{\mu}u_x)_x+\epsilon u^{\mu }u_x,\quad u = \varphi(t)\psi(x),\\
{\rm(c)}\quad u_{tt}=((e^u-1)^{-1}u_x)_x,\quad \ln(e^u-1)=\varphi(t)+\psi(x),\\
{\rm(d)}\quad u_{tt}=(e^u(e^u+1)^{-1}u_x)_x,\quad u-\ln(e^u+1)=\varphi(t)+\psi(x),\\
{\rm(e)}\quad u_{tt}=(e^u(e^u-1)^{-1}u_x)_x,\quad -u-\ln(e^u-1))=\varphi(t)+\psi(x),\\
{\rm(f)}\quad u_{tt}=((e^u+1)^{-1}u_x)_x,\quad \ln(e^u+1)=\varphi(t)+\psi(x),\\ 
{\rm(g)}\quad u_{tt}=(-\frac{p_i^2+ap_i+b}{h^2}u_x)_x+K(u)u_x,\quad
    \int^h\frac{dz}{p_i(z)}=\varphi(t)+\psi(x),\quad i=1,2,3,
\end{gather*}
where $\epsilon=0,1$, $\mu\ne0$, $p_i$ are described above.
\end{theorem}
Using this theorem and additional equivalence transformations, we
can construct some exact solutions for equations~\eqref{eqConstCoefTelegraphEq} and~\eqref{eqVarCoefTelegraphEq}.
For example, equation (a) in theorem~\ref{TheoremOnExpresOfEquWithFSV}
admits the functional separation solution $u = \varphi(t) + \psi(x)$,
where $\varphi$ and $\psi$ satisfy
the system
\[
\varphi''=\lambda e^{\varphi},\quad (e^{\psi}\psi')'+\epsilon e^{\psi}\psi'-\lambda=0,
\]
which can be solved explicitly by
\[
\varphi(t)=\ln\Big|\frac{c_1}{2\lambda}\sec\Big(\frac{\sqrt{c_1}}{2}t
+c_2\frac{\sqrt{c_1}}{2}\Big)^2\Big|,\quad
\psi(x)=\ln|\lambda x+c_3e^{-x}-c_4-{\lambda}|,\quad \epsilon=1.
\]
If we take $\epsilon=0$, the equation (a) is exactly Eq.~\eqref{eqWaveEqexp}
which admits the separation solution
\[
u=\ln\Big|\frac{\sec(\frac{1}{2c_1}t+\frac{c_2}{2c_1})^2}{2\lambda c_1^2}\Big|
+\ln\Big|\frac{\lambda}{2}x^2-c_3x+c_4\Big|.
\]
This solution is novel which can not be obtained classical
Lie--Ovsiannikov algorithm.  By using the transformation 3.4, we
obtain an interesting functional separation solution of the
equation~\eqref{EqVarCoeExpWave} as follows:
\[
u=\ln\Big|\frac{\sec(\frac{1}{2c_1}t\sign(\gamma+e^{-x})+\frac{c_2}{2c_1})^2}{2\lambda
c_1^2}\Big|+\ln\Big|\frac{\lambda}{2(\gamma+e^{-x})}-c_3+c_4(\gamma+e^{-x})\Big|.
\]

Equation (b) in theorem \ref{TheoremOnExpresOfEquWithFSV} admits
the separation solution $u = \varphi(t)\psi(x)$, where $\varphi$ and
$\psi$ satisfy the system
\[
\varphi''=\lambda
\varphi^{\mu+1},\quad (\psi^{\mu}\psi')'+\epsilon\psi^{\mu}\psi'-\lambda\psi=0.
\]
When taking $\mu=-1$, $\epsilon=0$, equation (b) become~\eqref{eqWaveEqpower-1}
which admits a new separation solution
\[
u=\frac{1}{2\lambda c_1^2}
\sec\Big(\frac{1}{2c_1}x^2+\frac{c_2}{2c_1}\Big)^2\Big(\frac{\lambda}{2}t^2+c_3t+c_4\Big).
\]
Using transformation 4.5, we can get a solution for
Eq.~\eqref{eqVarCoefWaveEqpower-1}
\[
u=e^{\gamma e^{-x}}\frac{1}{2\lambda
c_1^2}\sec\Big(\frac{1}{2c_1}e^{-2x}+\frac{c_2}{2c_1}\Big)^2\Big(\frac{\lambda}{2}t^2+c_3t+c_4\Big).
\]
Similarly, equation (c) admits the functionally separation
solution $u=\ln(1+\varphi(t)\psi(x))$, $\varphi$ and $\psi$ satisfy the
system
\[
\varphi'^2=\alpha \varphi^2-2\lambda
\varphi-\beta,\quad \psi'^2-\beta\psi^{4}+2\lambda\psi^3+\alpha
\psi^2=0.
\]
The implicit solution of this system is given by
\[
\int^{\varphi(t)}\frac{dz}{\sqrt{\alpha z^2-2\lambda z-\beta}}=t,\quad
\int^{\psi(x)}\frac{dy}{\sqrt{\beta y^4-2\lambda y^3-\alpha y^2}}=x.
\]
Equations (d), (e), (f) also admit the functionally
separation solution $u=-\ln(\varphi(t)\psi(x)-1)$,
$-\ln(1-\varphi(t)\psi(x))$, $\ln(\varphi(t)\psi(x)-1)$
respectively, $\varphi$ and $\psi$ satisfy the system
\[
\varphi'^2=\alpha \varphi^2-2\lambda \varphi-\beta,\quad
\psi'^2+\beta\psi^{4}+2\lambda\psi^3-\alpha \psi^2=0.\]
The implicit solution of this system is given by
\[
\int^{\varphi(t)}\frac{dz}{\sqrt{\alpha z^2-2\lambda z-\beta}}=t,\quad
\int^{\psi(x)}\frac{dy}{\sqrt{-\beta y^4-2\lambda y^3+\alpha y^2}}=x.
\]

\section{Conservation laws}\label{SectionOnConsLaws}

In this section we classify local conservation laws of equations~\eqref{eqVarCoefTelegraphEq}
with characteristics depending, at mostly, on $t$, $x$ and~$u$.
For classification we use the direct method described in~\cite{Popovych&Ivanova2004ConsLaws}.

To begin with, we adduce a necessary theoretical background on conservation laws,
following, e.g.,~\cite{Olver1986,Popovych&Ivanova2004ConsLaws} and
considering for simplicity the case of two independent variables~$t$ and~$x$.
See the above references for the general case.

Let~$\mathcal{L}$ be a system~$L(t,x,u_{(\rho)})=0$ of $l$ PDEs $L^1=0$, \ldots, $L^l=0$
for the unknown functions $u=(u^1,\ldots,u^m)$
of the independent variables~$t$ and~$x$.
Here $u_{(\rho)}$ denotes the set of all the partial derivatives of the functions $u$
of order not greater than~$\rho$, including $u$ as the derivatives of the zero order.

First we give an empiric definition of conservation laws.
A {\em conservation law} of the system~$\mathcal{L}$ is a divergence expression
\begin{equation}\label{conslaw}
D_tT(t,x,u_{(r)})+D_xX(t,x,u_{(r)})=0
\end{equation}
which vanishes for all solutions of~$\mathcal{L}$.
Here $D_t$ and $D_x$ are the operators of total differentiation with respect to $t$ and $x$, respectively.
The differential functions $T$ and $X$ are correspondingly called a {\em density} and a {\em flux}
of the conservation law and
the tuple $(T,X)$ is a \emph{conserved vector} of the conservation law.

The crucial notion of the theory of conservation laws is one of equivalence and triviality of conservation laws.
\begin{definition}\label{DefinitionOfConsVectorEquivalence}
Two conserved vectors $(T,X)$ and $(T',X')$ are {\em equivalent} if
there exist functions~$\hat T$, $\hat X$ and~$H$ of~$t$, $x$ and derivatives of~$u$ such that
$\hat T$ and $\hat X$ vanish for all solutions of~$\mathcal{L}$~and
$T'=T+\hat T+D_xH$, $X'=X+\hat X-D_tH$.
A conserved vector is called {\em trivial} if it is equivalent to the zeroth vector.
\end{definition}

The notion of linear dependence of conserved vectors is introduced in a similar way.
Namely, a set of conserved vectors is {\em linearly dependent}
iff a linear combination of them is a trivial conserved vector.

Conservation laws can be investigated in the above empiric framework.
However, for deeper understanding of the problem and absolutely correct calculations
a more rigorous definition of conservation laws should be used.

For any system~$\mathcal{L}$ of differential equations the set~$\CV(\mathcal{L})$ of conserved vectors of
its conservation laws is a linear space,
and the subset~$\CV_0(\mathcal{L})$ of trivial conserved vectors is a linear subspace in~$\CV(\mathcal{L})$.
The factor space~$\CL(\mathcal{L})=\CV(\mathcal{L})/\CV_0(\mathcal{L})$
coincides with the set of equivalence classes of~$\CV(\mathcal{L})$ with respect to the equivalence relation adduced in
definition~\ref{DefinitionOfConsVectorEquivalence}.

\begin{definition}\label{DefinitionOfConsLaws}
The elements of~$\CL(\mathcal{L})$ are called {\em conservation laws} of the system~$\mathcal{L}$,
and the whole factor space~$\CL(\mathcal{L})$ is called {\em the space of conservation laws} of~$\mathcal{L}$.
\end{definition}

That is why description of the set of conservation laws can be assumed
as finding~$\CL(\mathcal{L})$ that is equivalent to construction of either a basis if
$\dim \CL(\mathcal{L})<\infty$ or a system of generatrices in the infinite dimensional case.
The elements of~$\CV(\mathcal{L})$ which belong to the same equivalence class giving a conservation law~${\cal F}$
are considered all as conserved vectors of this conservation law,
and we will additionally identify elements from~$\CL(\mathcal{L})$ with their representatives
in~$\CV(\mathcal{L})$.
For $(T,X)\in\CV(\mathcal{L})$ and ${\cal F}\in\CL(\mathcal{L})$
the notation~$(T,X)\in {\cal F}$ will denote that $(T,X)$ is a conserved vector corresponding
to the conservation law~${\cal F}$.
In contrast to the order $r_{(T,X)}$ of a conserved vector~$(T,X)$ as the maximal order of derivatives
explicitly appearing in
the differential functions $T$ and~$X$,
the {\em order of the conservation law}~$\cal F$
is called $\min\{r_{(T,X)}\,|\,(T,X)\in{\cal F}\}$.
Under linear dependence of conservation laws we understand linear dependence of them as elements of~$\CL(\mathcal{L})$.
Therefore, in the framework of ``representative'' approach
conservation laws of a system~$\mathcal{L}$ are considered as {\em linearly dependent} if
there exists linear combination of their representatives, which is a trivial conserved vector.

Let the system~$\cal L$ be totally nondegenerate~\cite{Olver1986}.
Then application of the Hadamard lemma to the definition of conservation law and integrating by parts imply that
the left hand side of any conservation law of~$\mathcal L$ can be always presented up to the equivalence relation
as a linear combination of left hand sides of independent equations from $\mathcal L$
with coefficients~$\lambda^\mu$ being functions of $t$, $x$ and derivatives of~$u$:
\begin{equation}\label{CharFormOfConsLaw}
D_tT+D_xX=\lambda^1 L^1+\dots+\lambda^l L^l.
\end{equation}

\begin{definition}\label{DefCharForm}
Formula~\eqref{CharFormOfConsLaw} and the $l$-tuple $\lambda=(\lambda^1,\ldots,\lambda^l)$
are called the {\it characteristic form} and the {\it characteristic}
of the conservation law~$D_tT+D_xX=0$ correspondingly.
\end{definition}

The characteristic~$\lambda$ is {\em trivial} if it vanishes for all solutions of $\cal L$.
Since $\cal L$ is nondegenerate, the characteristics~$\lambda$ and~$\tilde\lambda$ satisfy~\eqref{CharFormOfConsLaw}
for the same conserved vector~$(T,X)$ and, therefore, are called {\em equivalent}
iff $\lambda-\tilde\lambda$ is a trivial characteristic.
Similarly to conserved vectors, the set~$\Ch(\mathcal{L})$ of characteristics
corresponding to conservation laws of the system~$\cal L$ is a linear space,
and the subset~$\Ch_0(\mathcal{L})$ of trivial characteristics is a linear subspace in~$\Ch(\mathcal{L})$.
The factor space~$\Ch_{\rm f}(\mathcal{L})=\Ch(\mathcal{L})/\Ch_0(\mathcal{L})$
coincides with the set of equivalence classes of~$\Ch(\mathcal{L})$
with respect to the above characteristic equivalence relation.

An important property of the class of equations in the conserved form is that it is preserved
under any point transformation (see, e.g.,~\cite{Popovych&Ivanova2004ConsLaws}).

\begin{proposition}\label{PropEqTrCL}
A point transformation~$g$: $\tilde t=t^g(t,x,u)$, $\tilde x=x^g(t,x,u)$, $\tilde u=u^g(t,x,u)$
prolonged to derivatives of~$u$
transforms the equation $D_tT+D_xX=0$ to the equation \mbox{$D_tT^g+D_xX^g=0$}.
The transformed conserved vector~$(T^g,X^g)$ is determined
by the formula
\begin{gather*}
T^g(\tilde x,\tilde u_{(r)})=\frac{T(x,u_{(r)})D_t\tilde t+
X(x,u_{(r)})D_x\tilde t}{D_t\tilde t\,D_x\tilde x-D_x\tilde t\,D_t\tilde x},
\\[1ex]
X^g(\tilde x,\tilde u_{(r)})=\frac{T(x,u_{(r)})D_t\tilde x+
X(x,u_{(r)})D_x\tilde x}{D_t\tilde t\,D_x\tilde x-D_x\tilde t\,D_t\tilde x}.
\end{gather*}
\end{proposition}

\begin{remark}
In the case of one dependent variable ($m=1$) $g$ can be a contact transformation:
$\tilde t=t^g(t,x,u_{(1)})$, $\tilde x=x^g(t,x,u_{(1)})$, $\tilde u_{(1)}=u^g_{(1)}(t,x,u_{(1)})$.
Similar note is true for the below statement.
\end{remark}

\begin{proposition}\label{PropositionOnInducedMapping}
Any point transformation $g$ between systems~$\mathcal{L}$ and~$\tilde{\mathcal{L}}$
induces a linear one-to-one mapping $g_*$ from~$\CV(\mathcal{L})$ into~$\CV(\tilde{\mathcal{L}})$,
which maps $\CV_0(\mathcal{L})$ into~$\CV_0(\tilde{\mathcal{L}})$
and generates a linear one-to-one mapping $g_{\rm f}$ from~$\CL(\mathcal{L})$ into~$\CL(\tilde{\mathcal{L}})$.
\end{proposition}

In such way, if a point transformation connects two systems of partial differential equations,
then the same transformation maps the set of conservation laws  of the first system to the set of conservation laws
of the second system and the space of characteristics of conservation laws to the space of characteristics.
Therefore, a group of equivalence transformations of a class of systems establishes a one-to-one correspondence
between conservation laws of systems from the given class. So, we can consider a problem of investigation of
conservation laws with respect to the equivalence group of a class of systems
of differential equations
This problem can be investigated in the way that is similar to group classification in classes of systems of
differential equations. Namely, we construct firstly the conservation laws that are defined for
all values of the arbitrary elements. (The corresponding conserved vectors may depend on the
arbitrary elements.) Then we classify, with respect to the equivalence group, arbitrary elements
for each of that the system admits additional conservation laws.

For more detail and rigorous proof of the correctness
of the above definitions and statements see~\cite{Popovych&Ivanova2004ConsLaws}.

Using the most direct method described in~\cite{Popovych&Ivanova2004ConsLaws} we prove the following theorem.

\begin{theorem}
A complete list of $G^{\sim}$-inequivalent equations~\eqref{eqVarCoefTelegraphEq} having nontrivial
conservation laws with characteristics of the zeroth order is exhausted by ones given in table~\ref{TableCLsClas}.
\end{theorem}

\setcounter{tbn}{0}

\setcounter{clnumber}{0}
\begin{center}\footnotesize\renewcommand{\arraystretch}{1.15}
Table~\refstepcounter{table}\label{TableCLsClas}\thetable.
Conservation laws of equations~\eqref{eqVarCoefTelegraphEq} \\[1ex]
\begin{tabular}{|l|c|c|c|l|}
\hline
N & $H(u)$ & $K(u)$ & $f(x)$  & \hfil Basis conservation laws \\
\hline
\refstepcounter{tbn}\label{CaseCLsForAllHForAllKForAllF}\thetbn & $\forall$ & $\forall$ &  $\forall$ &
\refstepcounter{clnumber}CL$^{\theclnumber}$, \refstepcounter{clnumber}CL$^{\theclnumber}$ \\
\refstepcounter{tbn}\label{CaseCLsForAllHK0ForallF}\thetbn & $\forall$ & $0$ &  $\forall$ &
 CL$^1$,  CL$^2$, \refstepcounter{clnumber}CL$^{\theclnumber}$, \refstepcounter{clnumber}CL$^{\theclnumber}$\\
\refstepcounter{tbn}\label{CaseCLsForAllHKHForallF}\thetbn & $\forall$ & $H$ &  $\forall$ &
   CL$^1$,  CL$^2$, \refstepcounter{clnumber}CL$^{\theclnumber}$, \refstepcounter{clnumber}CL$^{\theclnumber}$\\
\refstepcounter{tbn}\label{CaseCLsForAllHK1F1}\thetbn & $\forall$ & $1$ &  $1$ &
 CL$^1$,  CL$^2$, \refstepcounter{clnumber}CL$^{\theclnumber}$, \refstepcounter{clnumber}CL$^{\theclnumber}$\\
\refstepcounter{tbn}\label{CaseCLsForAllHK1Fx-1}\thetbn & $\forall$ & $1$ &  $x^{-1}$ &
 CL$^1$,  CL$^2$, \refstepcounter{clnumber}CL$^{\theclnumber}$, \refstepcounter{clnumber}CL$^{\theclnumber}$ \\
\refstepcounter{tbn}\label{CaseCLsForAllHKH+kFexp}\thetbn & $\forall$ & $H+k$ &  $e^x$
   & CL$^1$,  CL$^2$, \refstepcounter{clnumber}CL$^{\theclnumber}$, \refstepcounter{clnumber}CL$^{\theclnumber}$ \\
\refstepcounter{tbn}\label{CaseCLsForAllHKH-k2Fexpfracexp+c}\thetbn & $\forall$ & $H-k^2$ & $(1+ce^{-x})^{-1}$ &
  CL$^1$,  CL$^2$, \refstepcounter{clnumber}CL$^{\theclnumber}$, \refstepcounter{clnumber}CL$^{\theclnumber}$ \\
\refstepcounter{tbn}\label{CaseCLsForAllHKH+k2Fexpfracexp+c}\thetbn & $\forall$ & $H+k^2$ &  $(1+ce^{-x})^{-1}$ &
 CL$^1$,  CL$^2$, \refstepcounter{clnumber}CL$^{\theclnumber}$, \refstepcounter{clnumber}CL$^{\theclnumber}$ \\
\hline
\end{tabular}
\end{center}
{\footnotesize
Here $c$ and $k$ are arbitrary constants, $k\ne0$ in cases~\ref{CaseCLsForAllHKH-k2Fexpfracexp+c}
and~\ref{CaseCLsForAllHKH+k2Fexpfracexp+c}.
}

\smallskip

The conserved densities $T$ and fluxes $X$ of the presented conservation laws have the following forms:
\\[1ex] \setcounter{clnumber}{0}
\refstepcounter{clnumber} CL$^{\theclnumber}$:\quad $fu_t$, $-(Hu_x+\int K)$;\\[0.5ex]
\refstepcounter{clnumber} CL$^{\theclnumber}$:\quad $f(tu_t-u)$, $-t(Hu_x+\int K)$;\\[0.5ex]
\refstepcounter{clnumber} CL$^{\theclnumber}$:\quad $xfu_t$, $-(xHu_x-\int H)$;\\[0.5ex]
\refstepcounter{clnumber} CL$^{\theclnumber}$:\quad $xf(tu_t-u)$, $-t(xHu_x-\int H)$;\\[0.5ex]
\refstepcounter{clnumber} CL$^{\theclnumber}$:\quad $e^xfu_t$, $-e^xHu_x$;\\[0.5ex]
\refstepcounter{clnumber} CL$^{\theclnumber}$:\quad $e^xf(tu_t-u)$, $-e^xtHu_x$;\\[0.5ex]
\refstepcounter{clnumber} CL$^{\theclnumber}$:\quad $(2x-t^2)u_t+2tu$, $-(2x-t^2)(Hu_x+u)+2\int H$;\\[0.5ex]
\refstepcounter{clnumber} CL$^{\theclnumber}$:\quad $(6tx-t^3)u_t-(6x-3t^2)u$, $-(6tx-t^3)(Hu_x+u)+6t\int H$;\\[0.5ex]
\refstepcounter{clnumber} CL$^{\theclnumber}$:\quad $\sin tu_t-\cos tu$, $-\sin t(xHu_x+xu-\int H)$;\\[0.5ex]
\refstepcounter{clnumber} CL$^{\theclnumber}$:\quad $\cos tu_t+\sin tu$, $-\cos t(xHu_x+xu-\int H)$;\\[0.5ex]
\refstepcounter{clnumber} CL$^{\theclnumber}$:\quad
  $(2e^{2x}-kt^2e^x)u_t+2kte^xu$, $-(2e^{2x}-kt^2e^x)(Hu_x+ku)+kt^2e^x\int H$;\\[0.5ex]
\refstepcounter{clnumber} CL$^{\theclnumber}$:\quad
 $(6te^{2x}-kt^3e^x)u_t-(6e^2x-3kt^2e^x)u$, $-(6te^{2x}-kt^3e^x)(Hu_x+ku)+kt^3e^x\int H$;\\[0.5ex]
\refstepcounter{clnumber} CL$^{\theclnumber}$:\quad
$e^{x-kt}(u_t+ku)$, $-e^{-kt}(e^x+c)(Hu_x+u)-ce^{-kt}\int H$;\\[0.5ex]
\refstepcounter{clnumber} CL$^{\theclnumber}$:\quad
 $e^{x+kt}(u_t-ku)$, $-e^{kt}(e^x+c)(Hu_x+u)-ce^{kt}\int H$;\\[0.5ex]
\refstepcounter{clnumber} CL$^{\theclnumber}$:\quad
  $e^x(\sin ktu_t-k\cos ktu)$, $-\sin kt(e^x+c)(Hu_x+u)-c\sin kt\int H$;\\[0.5ex]
\refstepcounter{clnumber} CL$^{\theclnumber}$:\quad
 $e^x(\cos ktu_t+k\sin ktu)$, $-\cos kt(e^x+c)(Hu_x+u)-c\cos kt\int H$.\\ 

The above conservation laws can be used for construction of potential systems, potential symmetries
and potential conservation laws. We will present such analysis elsewhere.

\section{Conclusion}\label{SectionOnConclusion}

In summary, we have performed completely group classification of
the class of equations~\eqref{eqVarCoefTelegraphEq} by using the compatibility method and
additional equivalence transformations~\cite{Nikitin&Popovych2001,Popovych&Ivanova2004NVCDCEs}.
The main results
on classification are collected in tables~\ref{TableGrClasForAllH}--\ref{TableGrClasHpower} where we list
inequivalent cases of extensions with the corresponding Lie
invariance algebras.
Following the tables, we write down all the
additional equivalence transformations, reducing some equations
from our classification to others of simpler forms. For a number
of equations from the list of the reduced ones we construct
optimal systems of inequivalent subalgebras, corresponding Lie
ans\"{a}tze and exact solutions. By means of additional
equivalence transformations the solutions obtained are transformed
to the ones for the more interesting and complicated variable coefficient equations.
Functionally separation solutions are obtained for a number of equations
from class~\eqref{eqVarCoefTelegraphEq} via generalized conditional
symmetry method.

The present paper should be an inspiration for further
investigations of different properties of class~\eqref{eqVarCoefTelegraphEq}. For example,
one can classify the nonclassical (conditional) symmetries.
Furthermore, one can solve the general equivalence problem for any
pair of equations from class~\eqref{eqVarCoefTelegraphEq} with respect to the local
transformations so as to finding the group of all
possible local equivalence transformations (i.e. not only
continuous ones) in the whole class~\eqref{eqVarCoefTelegraphEq} as well as all the
conditional equivalence transformations.

Motivated by some elegant
results about potential symmetries of partial differential equations by Bluman
et.al~\cite{Bluman&Cheviakov&Ivanova2005,Bluman&Kumei1987,Bluman&Kumei1989,Bluman&Temuerchaolu&Sahadevan2005,
Bluman&Temuerchaolu2005a,Bluman&Temuerchaolu2005b}
and results of section~\ref{SectionOnConsLaws}, we also intend to
systematically calculate nonlocal symmetries and higher order local and potential conservation
laws, construct invariant and nonclassical solutions, as well as
obtain linearizations, etc, by investigating nonlocally related
potential systems and subsystems of variable coefficient equations~\eqref{eqVarCoefTelegraphEq}.
Further work along these lines would be extremely
interesting.

\subsection*{Acknowledgements}
Research of D-j.H. was partially supported by the National Key Basic Research Project of China
under the Grant NO.2004CB318000.

\end{document}